\documentclass[pre]{revtex4}
\usepackage{graphicx}
\usepackage{./fig}
\usepackage{times}
\usepackage{amssymb}

\begin{document}
\newcommand{\rme}{\mathrm{e}}
\newcommand{\rmd}{\mathrm{d}} 
\newcommand{\nn}{\nonumber}
\newcommand{\E}{\epsilon}
\arraycolsep0.5mm

\def\asinh{{\rm asinh \ }}
\def\atan{{\rm atan \ }}

% Allow for more pictures on a page
\renewcommand{\textfraction}{0.0}
\renewcommand{\topfraction}{1.0}
\renewcommand{\bottomfraction}{1.0}
\newlength{\diags}
\setlength{\diags}{0.205\columnwidth}

\evensidemargin -29pt %for letter -20pt
\oddsidemargin  -29pt %for letter -20pt

\title{\Large \bfseries \sffamily Wetting and Minimal Surfaces}

\author{\bf  \sffamily Constantin Bachas${}^{1,2}$, 
 Pierre Le Doussal${}^{1}$,
Kay J\"org Wiese${}^{1}$ \smallskip\smallskip\smallskip}
 
\affiliation{$\ ^1$ CNRS-Laboratoire de Physique Th{\'e}orique de
l'Ecole Normale Sup{\'e}rieure, 24 rue Lhomond, 75231 Cedex 05, Paris,
France. \\ 
\vskip -1.4mm $\ ^2$ Institut f\"ur Theoretische Physik,
ETH Z\"urich, 8093 Z\"urich, Switzerland.  \smallskip}

\date{\today \smallskip}

\begin{abstract}
We study minimal surfaces which arise in wetting and capillarity
phenomena.  Using conformal coordinates, we reduce the problem to a
set of coupled boundary equations for the contact line of the fluid
surface, and then derive simple diagrammatic rules to calculate the
non-linear corrections to the Joanny-de Gennes energy. We argue that
perturbation theory is quasi-local, i.e.\ that all geometric length
scales of the fluid container decouple from the short-wavelength
deformations of the contact line.  This is illustrated by a
calculation of the linearized interaction between contact lines on two
opposite parallel walls.  We present a simple algorithm to compute
the minimal surface and its energy based on 
these ideas. We also point out the intriguing
singularities that arise in the Legendre transformation from the pure
Dirichlet to the mixed Dirichlet-Neumann problem. 
\end{abstract}

\maketitle

\begin{widetext}

%\vskip 3mm

\section{Introduction}
\label{s:1}

Minimal surfaces, i.e.\ surfaces of minimal area with specified
boundary conditions, are found in many areas of physics, mathematics
and biology.  Their existence, uniqueness and other properties (such
as possible singularities or stability) are still actively studied by
mathematicians \cite{math}.  In the laboratory, minimal surfaces are
most commonly realized as soap films bounded by a given wire-frame, a
problem discussed already in 1873 by Plateau \cite{Plateau}.  In some
cases their morphology and stability have actually been elucidated
experimentally in this context \cite{Isenberg}.  Other systems where
minimal surfaces play a role include lipid-water solutions, diblock
copolymers, crystallography, protein structure, or liquid crystals
such as smectics \cite{boudaoud}.  They also arise as world-sheet
instantons in string theory, for example in the semiclassical,
fixed-angle high-energy limit of scattering amplitudes \cite{bp}.

\vskip 0.5mm The minimal surfaces that will interest us here arise in
the problem of { partial wetting} of a solid by a liquid
\cite{Gennes}.  In the standard experimental situation, a liquid with
free surface of area ${\cal A}$ (liquid-air interface) wets a flat
solid plane over an area ${\cal A}'$ (liquid-solid interface). The
free surface meets the solid plane along a line, called {\it contact
line}, at an angle $\theta$ which is defined locally.  The interfacial
energy is the difference ${\cal E} = \gamma {\cal A} - \gamma' {\cal
A}'$, where $\gamma$ is the energy per unit area (or surface tension)
of the liquid-air interface, and $\gamma'= \gamma_{SA} - \gamma_{SL}$
is the difference in surface tension between the solid-air (SA) and
the solid-liquid (SL) interfaces.  The force per unit length pushing a
segment of the contact line towards the unwetted region is thus $f = -
\gamma \cos \theta + \gamma'$. Requiring that it vanishes gives
Young's \cite{Young,Laplace} local equilibrium condition, $
\theta= {\rm arccos} (\gamma'/\gamma)$.  The minimal-surface problem
at hand is thus a problem with {\it mixed Neumann and Dirichlet }
boundary conditions.  In the idealized setting of an infinite liquid
container and a perfectly homogeneous planar wall, there exists a
simple solution to this problem: it is a planar liquid-air interface
meeting the wall along a straight contact line.  Strictly-speaking, as
we will discuss in section \ref{s:2}, the properties of the container
at infinity must be carefully chosen in order not to destabilize this
solution.  \vskip 0.5mm
 
Two extra forces play in fact a role in the general formulation of the
wetting problem.  The first comes from the drop in pressure across the
liquid-air interface, which adds to the Gibbs energy a volume term:
${\cal E} = \gamma {\cal A} - \gamma' {\cal A}' - p V$. Here $p$ is
the pressure difference and $V$ the volume of the fluid.  The free
surfaces that minimize this energy have constant rather than vanishing
mean curvature \cite{Laplace}. It is quite remarkable that the
corresponding equations are (at least formally) integrable, see for
instance \cite{helein}.  Note that in the special case of an
incompressible fluid, $p$ is a Lagrange multiplier determined by the
constraint that the ``droplet'' volume $V$ be fixed.  The second force
that plays in general a role is gravity, which introduces an
additional scale, the capillary length $\kappa^{-1} =(\gamma/\rho
g)^{1/2}$.  Here $\rho$ is the fluid mass density, and $g$ is the
gravitational acceleration.  In this paper we will study situations where both
pressure and gravity can be ignored. This is usually valid if one 
concentrates on length scales $\ll
\kappa^{-1}$, and considers a fluid connected to an
infinite reservoir so that effectively $p\simeq 0$.  Note that the
capillary length is typically of the order of a few millimeters, but
it can be made much larger in free-fall (e.g. space-based)
experiments, or if one replaces the air by a second non-mixing fluid
of roughly equal mass density.  Thus setting $\kappa \simeq p \simeq
0$ is a good approximation in a wide range of experimentally-feasible
situations, and we will do so in this paper. Technically, one can further
justify that gravity be ignored {\it at all scales} if a condition, identified 
below, is satisfied. 

\vskip 0.5mm 
What is in fact more questionable is the assumption of a
perfectly homogeneous wall. Indeed, in most of the experimental setups
of wetting, roughness and impurities of the solid substrate couple
directly to the position of the contact line, which may as a result be
effectively pinned.  Computing the energy of a {\it deformed} contact
line is thus a question of foremost importance.  For small
deformations, as Joanny and de Gennes (JdG) have shown
\cite{JoannyDeGennes1984}, the contact line obeys {\it non-local
linear elasticity}.  These linear equations may become unstable at
wavelengths comparable to some global-geometry scale, as several
earlier studies have established \cite{japonais}.  The issue of
non-linear elasticity, which becomes relevant for larger deformations,
has been addressed only recently \cite{GolestanianRaphael2001}.  It
could play a role \cite{LeDoussalWieseRaphaelGolestanian2004} in
resolving the apparent disagreement between recent experimental
measurements of contact-line roughness
\cite{PrevostRolleyGuthmann2002}, and renormalization-group
calculations near the depinning transition
\cite{LeDoussalWieseChauve2002} or numerical simulations
\cite{RossoKrauth2002} that were based on the JdG linear theory
\cite{MoulinetRossoKrauthRolley2004}.  To be sure, hysteresis and
other dynamical phenomena, which have attracted much of the recent
attention \cite{Pomeau}, may also prove important in interpreting the above
experimental data.  Nevertheless, a systematic analysis should start
with a thorough understanding of the non-linear and possibly non-local
effects in the simpler, equilibrium situation. This is the problem
that we will study here. 
  
\vskip 0.5mm  
 
The area of a minimal surface bounded by a given (closed) curve is
simple when expressed in conformal coordinates.  Non-linearities arise
because this choice of coordinates depends non-trivially on the
boundary curve, through the conformal-gauge (or Virasoro) conditions.
In this paper we develop systematic methods for solving the
ensuing non-local and non-linear equations, either in perturbation
theory or numerically. We focus, in particular, on the case of a
planar wall, and derive simple diagrammatic rules that calculate the
energy of a deformed contact line to any given order in the
deformation amplitude.  The method can be extended to more complicated
container geometries, but the details become more involved.  As a
relatively simple illustration, we show how to extend the rules,
and calculate the JdG linear theory, in the case of two contact lines
lying on parallel opposite walls.  We also describe a novel algorithm 
which finds the minimal surface energy with no need for surface triangulation. 
Finally, we discuss some general
properties of these perturbative expansions, which bear a fascinating
similarity to problems encountered in perturbative string theory.  We
hope to return to some of these questions, as well as to the
implications of our results for the wetting problem, in a future
publication.

\vskip 0.5mm  

The paper is organized as follows: In section \ref{s:2} we describe
our basic model, point out the need for global tadpole cancellation,
and discuss the relation of the mixed Neumann/Dirichlet to the pure
Dirichlet problem.  In section \ref{s:3} we give the formal solution
of the latter problem, for an arbitrary boundary curve, in terms of
conformal coordinates.  This is standard material which is included
here for completeness.  In section \ref{s:4} we specialize to the case
of a planar wall, derive the corresponding non-linear boundary
equations, and express the energy in terms of their solution.  We pay
particular attention to the decoupling of the large-volume cutoff,
which as we will explain is rather subtle.  In section \ref{s:5} we
solve the boundary equations perturbatively, and compute the
corrections to the JdG energy, up to quartic order.  Section \ref{s:6}
describes an alternative approach, using Lagrange-multiplier fields
and leading to a simple diagrammatic representation of the
perturbative expansion.  The numerical algorithm is presented in Section
\ref{s:C}. In section \ref{s:7} we extend this to the case of
two parallel walls, and calculate the quadratic interaction of the
contact lines.  Finally, in section \ref{s:8} we establish the finiteness of
the perturbative expansion order by order, and point out some
intriguing directions for future work. The Weierstrass parameterization
of our fluid surfaces, a calculation confirming the decoupling of the
large-volume cutoff are described,
respectively, in appendices  \ref{s:Weierstrass} and \ref{s:B}.

%\vskip 5mm      
  
%%%%%%%%%%%%%%%%%%%%%%%%%%%%%%%%%%%%%%%%  
 
\section{The Model}
\label{s:2}

We consider a fluid inside a tubular container $\Omega \times
\mathbb{R}$, where $\mathbb{R}$ corresponds to the height coordinate
$z$, and $\Omega$ is some (a priori arbitrary) connected region in the
$(x,y)$ plane, with boundary $\partial \Omega$.  Let us for now assume
that the fluid surface has no overhangs -- it can then be parameterized
by the height function $z(x,y)$.  We may express the energy functional as the following sum of
two-dimensional bulk and boundary terms:
\begin{equation}\label{grl}
{\cal E} = {\cal E}_{\rm bulk} + {\cal E}_{\rm bnry}  =
\int_{\Omega}\, \rmd x\, \rmd y \, \left( \gamma \sqrt{1 + (\partial_x
z)^2 \, +\, (\partial_y z)^2} \ - \ p z + \frac{1}{2} \rho g z^2 \right)  - 
\int_{\partial\Omega} \rmd l \, \gamma^\prime (l) \, z \ \ ,
\end{equation}
where $\rmd l$ is the infinitesimal length along the boundary of
$\Omega$.  The first term in (\ref{grl}) is the fluid-air interfacial
energy $\gamma {\cal A}$, the second is due to the difference in
pressure between air and fluid, the third to gravity, while the last
comes from the fluid-solid interface. For convenience, we have
slightly generalized the model so that the tension of this interface
may vary along the container walls, as can  be done by design. The
more general case of a $\gamma^\prime$ depending on both $l$ and $z$,
due for instance to the presence of impurities, will be discussed below. For now
$\gamma^\prime$ is only a function of $l$.

In the absence of gravity $g=0$, the minimum of the energy ${\cal E}$ is a surface of constant mean
curvature, with specified contact angles:
\begin{equation}\label{neum}
\vec \nabla \cdot \left( {\vec \nabla z\over \sqrt{1+
\vert \vec \nabla z\vert^2} } \right) = - { p\over \gamma }\ \ \ \ \ \ 
{\rm and}\ \ \ \ \ \ \  {\widehat n \cdot\vec \nabla z\over \sqrt{1+
\vert \vec \nabla z\vert^2}}  \Biggl\vert_{\partial\Omega}\,  = \,  \cos\theta (l) \, = \, 
{\gamma^\prime(l)\over \gamma}\ , 
\end{equation}
where $\vec \nabla = (\partial_x , \partial_y)$ and $\widehat n$ is a
unit vector normal to the boundary $\partial\Omega$. These non-linear
equations do not always admit a global solution, see e.g.\ \cite{Finn}.
A necessary (but not sufficient) condition for a solution to exist is
\begin{equation}\label{tadp}
Q = {p} \times {\rm Area}(\Omega) \, +\, \oint_{\partial\Omega} \rmd l \,
\gamma^\prime (l) = 0 \ .
\end{equation}
This is a condition of {\it average-force cancellation}: indeed, the
left-hand side of the above equation couples linearly to the zero mode
of $z(x,y)$, and would lead to a runaway solution if it did not
vanish, the energy being unbounded in that case. By analogy with string theory we may refer to
this as a global tadpole cancellation condition.
Note, in particular, that for a homogeneous wall, for which
$\gamma^\prime$ is constant, one must fine tune the ratio of perimeter
to area so that it equals $p/\gamma^\prime$.  If the average-force
condition is satisfied, the average height of the fluid surface
becomes a free dynamical parameter of the solution, analogous to
the string-theoretic {\it moduli}. Its role must be
examined with care as it threatens a priori the stability of any
perturbative expansion at weak disorder, and may thus lead to
qualitatively new behavior.  \vskip 0.5mm

The emergence of condition (\ref{tadp}) clearly originates from the
neglect of gravity. If $g \neq 0$, it is easy to see that the energy
is always bounded from below and that the fluid will tend to rise
such that $\int \rmd x\, \rmd y\, z \simeq Q/\rho g$, the well known
capillarity effect. Hence if $Q$ is non zero, one expects that the
theory studied here, obtained setting $g=0$, breaks down for
wavevectors $q < \kappa$ (hence especially for the zero
mode). However, the interesting point, discussed below, is that if one
imposes $Q=0$ then one can safely set $g=0$ and obtain a theory which
is well defined at all scales. This is the theory studied here. It is
illustrated in Fig.~\ref{f:compcoord}.

Let us consider minimizing the energy in two steps: We first solve the
bulk equations keeping the contact line fixed, i.e.\ we find the
surface of constant mean curvature, $z_h(x,y)$, such that the
restriction of $z_h$ to $\partial\Omega$ is a given function $h(l)$.
We denote the corresponding bulk energy (or {reduced energy
functional} in the language of \cite{japonais}) by $ E[h]:={\cal
E}_{\rm bulk}(z_h)$. The energy of the equilibrium configuration is
then the minimum over all contact lines of \begin{equation} E[h] -
\oint_{\partial\Omega}\, \gamma^\prime \, h\ .
\end{equation}
Thus $\gamma^\prime$ plays the role of a source, and the minimum
energy is just the {\it Legendre transform} of the reduced energy
functional. If $\gamma^\prime$ were to depend also on $z$, the source would
be field-dependent.  We will comment on the subtleties of this
Legendre transformation between the Dirichlet and Neumann problems in
the concluding section.

\begin{figure}
\includegraphics[width=0.65\textwidth]{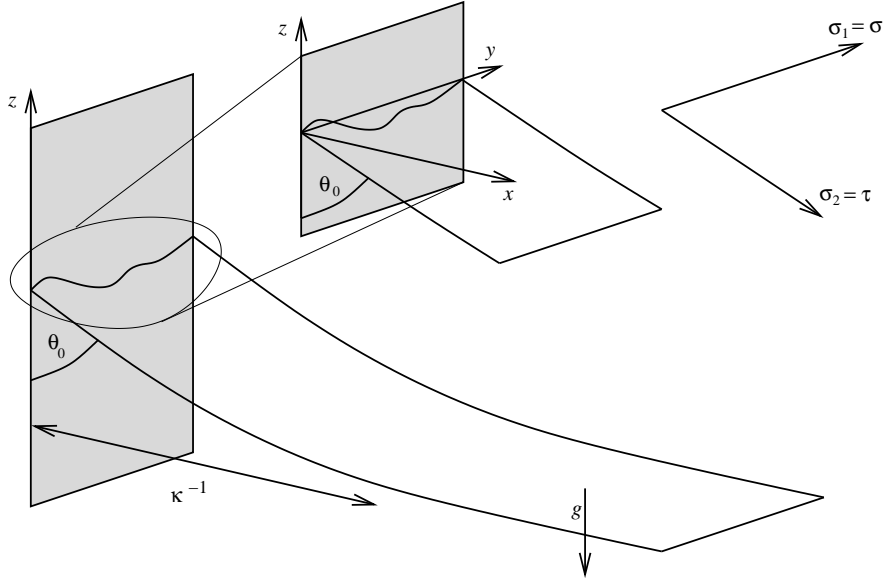} \caption{A
fluid surface bounded by a (shaded) planar wall, touching it at the
position of a (pinned) contact line.  At distances much larger than
the capillary length $1/\kappa$, it is flat and perpendicular to the
gravitational field (left).  Enlargement for distances smaller than
$1/\kappa$ (right) which is the range of scales studied here. The unperturbed 
surface is a plane, making an
angle $\theta_{0}$ with the wall. When perturbed, the conformal
parameters $(\sigma_1,\sigma_2)$ approach Cartesian coordinates far
from the wall, as discussed in section \ref{s:4}.}
\label{f:compcoord}
\end{figure}

\vskip 0.5mm
 
Let us describe the simplest configuration studied here, which
consists of a semi-infinite fluid bounded by a homogeneous planar wall
at $x=0$.  We assume from now on that $p=0$, and that the container at
infinity has been adjusted so that the global tadpole condition is
satisfied. The unperturbed fluid surface is then an inclined plane,
making a contact angle\ \ $\theta_0 = {\rm arccos}
(\gamma^\prime_0/\gamma)$ with the wall, as illustrated in figure
1. We are choosing the origin of coordinates so that the unperturbed
fluid surface intersects the wall along $z=0$, while the perturbed
contact line is given by $z = h(y)$. It turns out to be convenient for
the following to define:
\begin{equation}
 \tilde E[h] \, \equiv\,   E[h] - E[0]  -  \gamma\cos\theta_0 \int_y  h   \ , 
 \ 
\end{equation}
If the contact line deformations are concentrated in a finite region
one expects this energy difference to also be concentrated in a finite
region, and the outer boundaries of the container to decouple.  More
generally, the simple planar model of figure \ref{f:compcoord} should
give an adequate description of the physics if all other distance
scales of the system (including the capillary length, $\gamma/p$, and
all geometric scales) are much larger than the typical deformation
wavelength.  We will come back to this subtle issue later on. Note
that we have included in the energy difference the contribution,
${\cal E}_{\rm bnry}$, of the homogeneous wall. This means that
$\tilde E[h]$ should start out as a quadratic functional for small
$h(y)$.

Let us briefly mention the case of impurity disorder. In this case
translation symmetry of the tube is in general broken by the roughness
of the wall. The effect of impurities can then be modeled by a
variable fluid-solid tension, and the boundary term in (\ref{grl})
becomes:
\begin{equation}\label{spcase}
   {\cal E}_{\rm bnry} \, =\, - \oint_{\partial\Omega} \rmd l
\int_{0}^{z} \rmd \zeta\, \gamma^\prime (l, \zeta) \ .
\end{equation}
The two stage minimization can then be summarized as follows. One writes:
\begin{equation}
\gamma^\prime(l,z) = \gamma^\prime_0(l) 
+ {\Delta \gamma^\prime}(l, z) \ , 
\end{equation}
where $\gamma^\prime_0(l)=\gamma \cos \theta_0(l)$ is some average or
reference value, and defines the shifted functional:
\begin{equation}
 \tilde E[h] \, \equiv\, E[h] - E[0] - \gamma
\oint_{\partial\Omega}\rmd l\, \cos \theta_0(l) h(l)
\end{equation}
Because of disorder the impurities generate a potential for the zero
mode $z_0$ of $z(x,y)$ and the condition (\ref{tadp}) cannot hold in
general. However we can still impose this condition ``on average''
$\oint \rmd l\, \gamma^\prime_0(l) =0$ and compute the corresponding
$\tilde E[h]$. It is this functional which is studied here: it obeys
quasi-locality and is well defined for $g=0$. Once $\tilde E[h]$ is
known, finding the (equilibrium) position of the contact line amounts
to solving in the second stage of the minimization:
\begin{equation}
 \min_{h(l)} \left[ \tilde E[h] - \oint dl \int_0^{h(l)} d \zeta
\Delta \gamma^\prime (l, \zeta) \right] \ .
\end{equation}
This can be viewed as a generalized Legendre transformation,
which we will not study here. The aim of this paper being simply to
characterize $\tilde E[h]$ in presence of an average contact angle. We
will use expressions such as pinning condition, or pinned
configuration in the following only to denote the fixed-$h$
conditions.

%\vskip 3mm  

 %%%%%%%%%%%%%%%%%%%%%%%%%%%%%%%%%%%%%%%%%%%
 %%%%%%%%%%%%%%%%%%%%%%%%%%%%%%%%%%%%%%%%%%%

\section{Conformal coordinates} 
\label{s:3} 

Computing the area of a minimal surface bounded by a continuous closed
curve $\vec {\rm v}(s)$ is a classical problem of applied mathematics.
In this section we will explain how, in conformal gauge, it reduces to
a (non-linear and non-local) equation for a function of one variable
on the boundary.  Let $\vec r(\sigma_1,\sigma_2)$ be an arbitrary
parameterization of the surface, i.e.\ $\vec r = (x,y,z)$ is the
position of the surface $\Sigma$ corresponding to the values of the
two (a priori arbitrarily-chosen) parameters $(\sigma_1,\sigma_2)$.
We will assume that $\Sigma$ has the topology of a disk, and that the
parameterization is global, i.e.\ that there is a one-to-one
correspondence between points of $\Sigma$ and points in some parameter
domain ${\cal D}\subset {\mathbb{R}}^2$.  One should of course keep in
mind that, for some boundary curves, these assumptions may have to be
relaxed.  In terms of the induced metric $g_{ab} = \partial_{a} \vec r
\cdot \partial_{b} \vec r \ $, the area of $\Sigma$ reads:
\begin{equation}\label{area}
{\cal A} =  \int\hskip -2.1 mm \int_{\cal D} \rmd \sigma_{1}\rmd \sigma_{2}\, \sqrt{\det g}\ .  
\end{equation} 
This expression is invariant under any reparametrization with
non-vanishing Jacobian, i.e.\ $\sigma_1\to \tilde\sigma_1
(\sigma_1,\sigma_2)$ and $\sigma_2\to\tilde\sigma_2(
\sigma_1,\sigma_2)$ with ${\det}\, (\partial_a\tilde\sigma_b) \not=
0$.  For a surface without `overhangs' we may use this freedom to set
$(\sigma_1, \sigma_2)=(x,y)$, in which case (\ref{area}) reduces to
the expression for the area used in eq.~(\ref{grl}).  This is a
useful parameterization when $\partial_x z$ and $\partial_y z$ are
small, but more generally the minimization of the area in this gauge
leads to non-linear partial differential equations in two variables,
which are hard to solve.

\vskip 1mm 

A more convenient choice is  conformal coordinates,
which are defined implicitly by the two conditions~:
\begin{eqnarray}\label{conf1n}
\partial_{1} \vec r \cdot \partial_{2}\vec r =0  
\ \ \ {\rm and} \ \ \ \ \ 
\partial_{1}\vec r \cdot \partial_{1} \vec r = \partial_{2}\vec r
\cdot \partial_{2} \vec r\ .  
\end{eqnarray}
Put in words, the two vector fields tangent to the surface are
orthogonal everywhere and of equal, not necessarily constant, length.
[As the reader can easily verify, the parameterization
$(\sigma_1,\sigma_2)=(x,y)$ is conformal only in the special case of
constant $z$.]  It follows from (\ref{conf1n}) that $g_{ab} = \Phi^2\,
\delta_{ab}$, where $\Phi^2 = \partial_{1}\vec r \cdot \partial_{1}
\vec r $ is the so-called ``conformal factor''. Thus in this gauge the
area can be written as
\begin{equation}\label{NG1n}
{\cal A} = {1\over 2} \, \int\hskip -2.1 mm \int_{\cal D} \rmd
\sigma_{1}\rmd \sigma_{2}\, (\partial_{1}\vec r \cdot \partial_{1}
\vec r + \partial_{2}\vec r \cdot \partial_{2} \vec r) \ ,
\end{equation} 
and the variational equations are the Laplace equations in two
dimensions~:
\begin{equation}\label{j1}
  \partial_{a}\left( 
\sqrt{\det g}\,  g^{ab} \partial_{b}\,  \vec r \right) =
 \left(\partial_{1}^{2}+\partial_{2}^{2} \right) \vec r  = 0 \ . 
\end{equation} 
The embedding coordinates $(x,y,z)$ are therefore harmonic functions
of $(\sigma_1,\sigma_2)$, and can be written as the real parts of
analytic functions of the complex variable $w=
(\sigma_1+i\sigma_2)/2$~:
\begin{equation}
x(w,\bar w) = 2\, {\rm Re}\, X(w) \ \ ,  \ \ \ y(w,\bar w)  =  2\, {\rm Re}\, Y(w) \ \ , 
 \ \ \ z(w,\bar w)  = 2\, {\rm Re}\, Z(w) \ \ .
\end{equation} 
This property of harmonic functions is very special to two dimensions.
Our problem is now to determine $X$, $Y$ and $Z$ for the given
boundary curve $\vec {\rm v}(s)$.  \vskip 1mm

To this end, note first that if the surface is non-singular and
bounded, the functions $X, Y$ and $Z$ must be analytic in the interior
of the domain ${\cal D}$. They are furthermore related by the two
conformal-gauge conditions (\ref{conf1n}), which can be combined in
the following equivalent form~: \begin{equation}\label{conf4n}
\left(\partial_{1}- i \partial_{2} \right)\vec r \cdot
\left(\partial_{1}- i\partial_{2} \right) \vec r\ =\ (X^\prime)^{2}+
(Y')^{2} + (Z')^{2}\ = \ 0\ ,
\end{equation}
where the prime denotes differentiation with respect to $w$.  This
rewriting makes manifest the residual freedom of analytic
reparametrizations of $w$.  Such complex-analytic changes of
coordinates preserve indeed the conformal condition (\ref{conf4n}),
and can be used to map the parameter domain to any convenient
simply-connected region in $\mathbb{C}$.  Let us assume, for instance,
that ${\cal D} = \{ w\in \mathbb{C}, \vert w\vert\leq 1\}$ is the unit
disk.  We write $w=\rho e^{i\phi}$, and denote by $\vec r(\phi) :=
\vec r(\phi, \rho=1)$ the boundary curve parameterized by the special
conformal coordinate $\phi$.  Note that $\vec r(\phi)$ has a unique
harmonic extension to the interior of the disk, and thus determines
unambiguously the minimal surface.  This follows easily from the fact
that the analytic function $X(w)$ admits a Taylor expansion
\begin{equation} 
X(w) = \sum_{n=0}^\infty X_n w^n\ ,
\end{equation}
so that its restriction to the boundary has no negative-frequency
Fourier modes, when identifying $w^{n}=\rme^{i \phi n}$.  Thus, to
extend $x(\phi)$ to the interior of the disk, we need only split it
into positive- and negative-frequency parts, $x(\phi)= x_+(\phi)+
x_-(\phi)$.  Then $x_+$ can be extended to $X(w)$ by the replacement
$e^{i\phi}\to w$, while $x_- = \bar x_+$ extends to the
complex-conjugate anti-analytic function $\bar X(\bar w) = \overline{X(w)}$.  If
$x(\phi)$ has a zero mode, it must be split equally between the two
parts.  A simple calculation leads in fact to the following Cauchy
relation between $X(w)$ and the boundary restriction of $x$:
\begin{equation}\label{intt}
X(w) = {1\over 4\pi}\int_{0}^{2\pi} \rmd \phi' \, x(\phi ')\,
{e^{i\phi' }+w\over e^{i\phi'}-w} \ .
\end{equation}
Similar relations hold of course between $Y(w)$ and $y(\phi)$, and
also $Z(w)$ and $z(\phi)$.  It is, furthermore, easy to check that
since $X(e^{i\phi}) = x_+(\phi)$, the conformal-gauge condition
(\ref{conf4n}) is equivalent to
\begin{equation}\label{cgc}
{\rmd \vec r_+\over \rmd\phi}\cdot {\rmd \vec r_+\over \rmd\phi} = 0 \
\ \ \ \ {\rm for}\ {\rm all}\ \phi \in [0,2\pi]\ .
\end{equation}
  
Let us go back now to the expression (\ref{NG1n}) for the
area. If the surface is minimal, integrating by parts and using
Laplace's equation allows to rewrite its area as a boundary integral:
\begin{equation}\label{mina}
{\cal A}_{\rm min} = {1\over 2} \, \int_0^1\rho \rmd
\rho\int_{0}^{2\pi} \rmd\phi \, ( \partial_{\rho}\vec r \cdot
\partial_{\rho} \vec r + \rho^{-2} \partial_\phi\vec r\cdot
\partial_\phi\vec r)\ = \, {1\over 2} \int_{0}^{2\pi} \rmd\phi \ \vec
r \cdot \partial_\rho \vec r\ \Bigr\vert_{\rho =1}\ .
\end{equation}
The integrand involves the radial derivative of $\vec r$, but with the
help of Cauchy's equation ($\rho\partial_\rho X = -i\partial_\phi X$,
and similarly for the functions $Y$ and $Z$) we can convert this to an
angular derivative, with the result:
\begin{equation}\label{secV}
{\cal A}_{\rm min}\, =\, {i\over 2} \int_{0}^{2\pi} \rmd\phi \left(\vec r_+
\cdot {\rmd \vec r_-\over \rmd\phi} - \vec r_- \cdot {\rmd \vec
r_+\over\rmd\phi}\right)\ = \ 2\pi \sum_{n=1}^\infty n\, \vert \vec
r_n \vert^2 \ .
\end{equation}
Here $\vec r_n$ is the Fourier transform of the function on the circle
$\vec r (\phi) = \sum_n \vec r_n e^{i n \phi} $.  For later use, we also give two alternative
(equivalent) expressions for the minimal area:
\begin{equation}\label{minan}
{\cal A}_{\rm min}\, =\, -{1\over 4\pi} \int_\phi
\int_{\phi^\prime}\ {\rmd \vec r\over \rmd\phi}\cdot {\rmd \vec
r\over \rmd\phi^\prime}\ {\rm log}\, \sin^2\left({\phi -\phi^\prime\over
2}\right)\ = {1\over 16\pi} \int_\phi \int_{\phi^\prime}\ { \left \vert
\vec r(\phi ) - \vec r(\phi^\prime )\right \vert^2 \over \left( \sin
{1\over 2} {\left(\phi -\phi^\prime  \right)} \right)^2 } \
.
\end{equation}
The first can be obtained from eq.~(\ref{secV}) by Fourier transform,
while the second follows by a double integration by parts and the fact
that, thanks to the $i\epsilon$ prescription, only the cross term in
the numerator contributes.  Note that for suitably smooth $\vec
r(\phi)$ these integrals are manifestly finite in the
$\phi\to\phi^\prime$ region (hence the $i \epsilon$ can be dropped
in the final expression - but not if one expands the square).

\vskip 0.5mm 

We have thus succeeded to express the minimal area as an explicit
(non-local, but quadratic) functional of $\vec r(\phi)$, so one may
think that our problem is effectively solved.  This is, however, not
quite the case, because the transformation from the original parameter
of the boundary, to the special conformal coordinate $\phi$, depends
itself non-trivially on the boundary curve.  To make this relation
explicit, let us write $s = f(\phi)$, so that $\vec r(\phi) = \vec
{\rm v}(f(\phi))$. A straightforward calculation starting from the
integral expression (\ref{intt}) gives
\begin{equation}
 {\rmd \vec r_+ \over \rmd\phi} \ = \ -{i\over 8\pi}\int \rmd \phi^\prime
   \ {\vec {\rm v}(f(\phi^\prime)) \over 
 {\rm sin}^{2}({\phi-\phi^\prime +i\epsilon\over 2}) }  \ . 
\end{equation}
Plugging this in the gauge condition (\ref{cgc}) leads to a non-linear
integral equation, that can be used (in principle) to determine
$f(\phi)$ for any given boundary curve $\vec {\rm v}(s)$.  This is
still a non-trivial task, but we have at least reduced the
minimal-surface problem to one involving only one unknown function of
a single variable.  In some cases, the problem can be simplified
further by using the residual freedom of conformal transformations to
map the unit disk to a suitably-chosen domain.  Such is the case when
the contact line lies on a plane, as we will now see.

%%%%%%%%%%%%%%%%%%%%%%%%%%%%%%%%%%%%%%%%%%%%%

\section{Case of a planar wall}
\label{s:4}

\subsection{The boundary equations} In the configuration of fig.\,1
the contact line is restricted to a planar wall, located at $x=0$.
Assuming that it has no overhangs, such a contact line is naturally
parameterized by the height function $z=h(y)$. We want to adapt our
previous general discussion to this special situation. The story is
somewhat simplified by using the convenient conformal coordinates
(reminiscent of the proper-time gauge of string theory):
\begin{equation}\label{choice} X = -ic\, w = -{ic\over 2}\, (\sigma
+i\tau) \ , \ \ \ \ {\rm so\ that}\ \ \ \ x= 2 {\rm Re}\, X = c\,
\tau\ .
\end{equation} 
Here $c$ is a positive constant, and we have traded
$(\sigma_1,\sigma_2)$ for the lighter notation $(\sigma,\tau)$.  In
imposing (\ref{choice}) we have used the residual freedom of conformal
transformations, and the fact that $X$ is an analytic function.  Note,
however, that this choice of gauge might be obstructed globally, as we
will explain in appendix A.  Since the fluid surface extends out to
infinity, the new parameter domain is the upper-half complex plane,
${\cal D} = \{w\in \mathbb{C}, {\rm Im}w \geq 0\}$.  Later we will
consider a second wall at $x=L$, in which case ${\cal D}$ will be the
infinite strip $0\leq \tau\leq L/c$.  The points at infinity must
actually be treated with care: the right procedure is to first make
${\cal D}$ finite, by bounding the fluid with outer walls, then move
these outer boundaries to infinity.

\vskip 1mm We will be here interested in surfaces that approach
asymptotically the inclined plane
\begin{equation}\label{rnot}
\vec r_0 = (\sin\theta_0\, \, \tau, \, \sigma, \, -\cos\theta_0\, \tau) \ .  
\end{equation}
It is therefore convenient to choose $c =\sin\theta_0$, and to define
the difference
\begin{equation}
\Delta\vec r =  \vec r -  \vec r_0  \ , \ \ \ {\rm with} \ \ \ \ \Delta\vec r \equiv (0, \tilde y, \tilde z)
\ . 
\end{equation} 
Note that the gauge condition (\ref{choice}) ensures that the first
component of $\Delta\vec r$ is identically zero.  Since the components
of both $\vec r$ and $\vec r_0$ are harmonic, so are those of their
difference $\Delta\vec r$.  We can in fact write $\tilde y(w,\bar w) =
2{\rm Re}\, \tilde Y(w)$ and $\tilde z(w,\bar w) = 2{\rm Re}\, \tilde
Z(w)$, where the new analytic functions are given by
\begin{equation}\label{logic}
 \tilde Y = Y - w\ , \ \ \ {\rm and}\ \ \tilde Z = Z - i \cos\theta_0\, w\ . 
\end{equation} 
Following the same logic as in section \ref{s:3}, we also define the
restrictions of $\tilde y$ and $\tilde z$ to the real axis, $\tilde
y(\sigma) \equiv \tilde y(\sigma, \tau= 0)$ and $\tilde z(\sigma)
\equiv \tilde z(\sigma, \tau=0)$.  The extension of these functions to
the upper-half plane is uniquely determined by the property that they
should be both bounded and harmonic.  Indeed, the analytic function
$\tilde Y$ must have a Fourier-Laplace expansion involving only
positive-frequency modes:
\begin{equation}\label{Laplace}
 \tilde Y(w) = \int_{0}^\infty\frac{\rmd k}{2\pi }\, \, \tilde Y_k\,
e^{2ikw}\ \ \ \Longleftrightarrow \ \ \ \tilde y(\sigma) =
\int_{0}^\infty \frac{\rmd k}{2\pi}\, \, (\tilde Y_k\, e^{ik\sigma} +
c.c. )\ \ ,
\end{equation} 
since it would otherwise diverge when $\tau\to\infty$.  Thus, to
extend $\tilde y(\sigma)$ to the upper-half plane, we must first split
it into its positive- and negative-frequency parts, $\tilde y(\sigma)
= \tilde y_+(\sigma) + \tilde y_-(\sigma)$, then extend $\tilde y_+$
analytically and $\tilde y_-$ as its complex-conjugate anti-analytic
function.  The Cauchy integral formula relating $\tilde Y(w)$ and
$\tilde y(\sigma)$ reads \begin{equation}\label{invLaplace} \tilde
Y(w) = {i \over 2\pi} \int_{-\infty}^\infty \rmd\sigma \ {\tilde
y(\sigma)\over 2w-\sigma}\ \ .  \end{equation} The right-hand side is
analytic in the upper-half complex plane provided that $\tilde
y(\sigma)$ vanishes at infinity.  Of course a similar formula relates
also $\tilde z(\sigma)$ to its analytic counterpart $\tilde Z(w)$.

\vskip 1mm Our problem is thus reduced to that of finding the two real
functions on the real axis, $\tilde z(\sigma)$ and $\tilde y(\sigma)$.
These are related by the pinning condition of the contact line:
\begin{equation}\label{bc1n} \tilde z(\sigma) = h \Bigl(\sigma +
\tilde y(\sigma)\Bigr)\ \ .
\end{equation}
Furthermore, they must obey the conformal constraint
(\ref{conf4n}). After inserting the expressions (\ref{logic}), and
using the obvious identities $\tilde y_+(\sigma) = \tilde Y(\sigma/2)$
and $\tilde z_+(\sigma) = \tilde Z(\sigma/2)$, this constraint reads:
\begin{equation}\label{constr}
{\rmd \tilde y_+\over \rmd\sigma} + i \cos\theta_0 \, {\rmd \tilde
z_+\over \rmd\sigma} = - \left({\rmd \tilde y_+\over
\rmd\sigma}\right)^{\!\! 2} - \left({\rmd \tilde z_+\over
\rmd\sigma}\right)^{\!\!2}\ .
\end{equation}
The pair of coupled, non-local equations (\ref{bc1n}) and
(\ref{constr}) is in principle sufficient to determine $\tilde
z(\sigma)$ and $\tilde y(\sigma)$, and hence also the complete shape
of the fluid surface.  In the following sections we will discuss how
to solve these equations numerically, or by a series expansion in
powers of $h(y)$.  First, however, we must express the energy in terms
of the two boundary functions $\tilde z(\sigma)$ and $\tilde
y(\sigma)$.

%%%%%%%%%%%%%%%%%%%%%%%%%%%%%%%%%%%%%%%%%%%%

\subsection{Expression for the energy}  
\begin{figure}
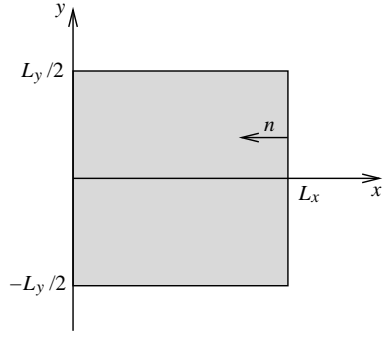

\fig{5cm}{inward}
\caption{The domain $C$ and the inward pointing normal.}
\label{normal}
\end{figure}
The area of an infinite fluid surface is, clearly, infinite. However,
for a localized deformation of the contact line, i.e.\ for\ \ $h(y)\to
0$ when $y\to\pm\infty$, we expect the difference in area,
$\tilde{\cal A}_{\rm min} \equiv {\cal A}_{\rm min}[h] - {\cal A}_{\rm
min}[0]$, to be finite.  To calculate this difference, we will
introduce as a physical cutoff a tubular container $C= \Omega\times
\mathbb{R}$, with $\Omega$ a rectangle of size $L_x\times L_y$ in the
$(x,y)$ plane. We define the associated characteristic function
\begin{equation}\label{contn} \Theta_C(\vec r) := \Theta(x) \,
\Theta \!\left(y+{L_y\over 2}\right) \Theta\!\left({L_y\over
2}-y\right) \Theta(L_x - x )\ = \cases{ 1 \ \ \ \ {\rm if} \ \ \vec
r\in C \cr 0 \ \ \ \ {\rm otherwise}\ , }
\end{equation}
with $\Theta(a)$ the usual Heaviside step function.  The difference of
the areas then reads:
\begin{eqnarray}\label{amin}
\tilde{\cal A}_{\rm min} = \ {1\over 2} \int\hskip -2.5mm\int_{\cal
\mathbb{R}\times\mathbb{R}} \ \Bigl[ \Theta_C(\vec r)\, \partial_a
\vec r \cdot \partial_b \vec r \, - \, \Theta_C(\vec r_0)\, \partial_a
\vec r_0 \cdot \partial_b \vec r_0\, \Bigr] \ \delta^{ab}\ ,
\end{eqnarray}
where, after evaluating the right-hand side, we should take the limit
$L_x, L_y\to\infty$.  Note that cutting off directly the parameter
range could give a wrong answer, because the same value of
$(\sigma,\tau)$ need not correspond to the same value of $(x,y)$ on
the planar and on the deformed surface.

\vskip 1mm
  
Expanding the integrand of eq.~(\ref{amin}) in powers of $\Delta\vec
r$, and using the fact that
\begin{equation}
\Theta_C(\vec r_0+\Delta\vec r) = 
\Theta_C(\vec r_0) +  \delta_C(\vec r_0) \, \hat n\cdot\Delta\vec r
+ \cdots \ , 
\end{equation}
where $\delta_C$ is the delta function localized on the boundary of
$C$ and $\hat n$ is the inward-pointing normal unit vector, leads to
the following expression for the area difference:
\begin{eqnarray}\label{j8n}
\tilde{\cal A}_{\rm min} = {1\over 2} \int\hskip -2mm \int_{\cal D} (
\partial_a \Delta \vec r \cdot \partial_b \Delta\vec r + 2 \partial_a
\Delta \vec r\cdot \partial_b \vec r_0) \ \delta^{ab}\ +
\int_{\partial{\cal D}} \vert \hat n\cdot \partial_\perp \vec
r_0\vert^{-1} \ \hat n \cdot \Delta\vec r + \cdots \ \ .
\end{eqnarray}
Here ${\cal D} = [0, L_x/\sin\theta_0]\times [-L_y/2, L_y/2]$ is the
parameter domain defined by the condition $\Theta_C(\vec r_0) = 1$,
and $\partial\cal D$ is its boundary.  The last term in the above
equation accounts for the fact that the cutoff corresponds to a
container in physical space, rather than in the space of parameters
$(\sigma,\tau)$.  The factor $\vert \hat n\cdot \partial_\perp \vec
r_0\vert^{-1}$, with $\partial_\perp:= \hat n\cdot \vec \partial $ a
derivative in the direction normal to ${\partial D}$, is the Jacobian
that arises upon converting $\delta_C(\vec r_0)$ to a
$\delta$-function in parameter space.  The neglected terms involve
higher powers of $\hat n\cdot \Delta\vec r$, and one or more partial
derivatives.  They vanish on the outer boundary, provided $\vec \Delta
r\to 0$ at infinity, and on the $x=0$ wall where $\hat n\cdot\vec
\Delta r=0$ for our choice of gauge.  Note that in deriving expression
(\ref{j8n}) we used the equality $\partial_a \vec r_0\cdot \partial_b
\vec r_0\, \delta^{ab} = 2$, which follows easily from (\ref{rnot}).
 
\vskip 0.5mm  

Using Stoke's theorem and Laplace's equation we can express all the
terms in (\ref{j8n}) as boundary integrals,
\begin{eqnarray}\label{bpi2n} \tilde{\cal A}_{\rm min} = \ -
\int_{\partial D} \Bigl( {1\over 2} \Delta\vec r \cdot \partial_\perp
\Delta \vec r \ + \Delta\vec r \cdot \partial_\perp \vec r_0 \ - \vert
\hat n\cdot \partial_\perp \vec r_0\vert^{-1}\ \hat n \cdot \Delta\vec
r \Bigr)\ .
\end{eqnarray}
Let us consider first the $\sigma=\pm L_y/2$ boundaries: since $\hat
n = \mp (0, 1, 0)$ there, the last two terms cancel exactly one
another, while the term quadratic in $\Delta\vec r$ does not
contribute as long as $\Delta\vec r\to 0$ at infinity.  This  term
does not contribute, for the same reason, at the $\tau =
L_x/\sin\theta_0$ boundary.  Finally, at both $\tau =0$ and $\tau =
L_x/\sin\theta_0$ we have $\hat n \cdot \Delta\vec r =0$, since $\hat
n = \pm (1, 0, 0)$ on these boundaries and, with our choice of gauge,
$\Delta\vec r = (0, \tilde y, \tilde z)$. Putting all these facts
together we obtain: 
\begin{eqnarray}\label{bpi3n} \tilde{\cal A}_{\rm
min} = \ \int_{-\infty}^\infty \rmd{\sigma} \ \Bigl\{ -{1\over 2} (
\tilde y\,\partial_\tau\tilde y + \tilde z\,\partial_\tau\tilde
z)\Bigr\vert_{\tau=0} \ +\cos\theta_0 \, \tilde z\Bigr\vert_{\tau =0}
- \cos\theta_0 \, \tilde z\Bigr\vert_{\tau =L_x/\sin\theta_0}\ \Bigr\}
\ ,
\end{eqnarray}
where the limit $L_y\to\infty$ has already been taken on the
right-hand-side. The $\tau$-derivatives in the first term can be
converted to $\sigma$-derivatives with the help of the Cauchy
equation.  As for the last two linear terms, they cancel because
$\tilde z$ is harmonic (both are proportional to the same $k=0$ Fourier mode).
Thus the difference of the areas reads:
\begin{eqnarray}\label{misle}
\tilde{\cal A}_{\rm min} =  
 \ {1\over 2} \int_{-\infty}^\infty \rmd\sigma \,  
  \Bigl[  \, i \tilde y_+ {\rmd \tilde y_-\over \rmd\sigma}  
  + i \tilde z_+ {\rmd \tilde z_-\over \rmd\sigma}   
\,  +\,  c.c. \Bigr]     \ .   
\end{eqnarray}
\vskip 1mm

Although this calculation is correct, the cancellation of the linear
terms is, from the physical point of view, rather misleading.  It
involves two opposite walls which are infinitely far apart in the
$L_x\to\infty$ limit, and looks therefore highly non-local.  A
physically more significant cancellation occurs in the energy
functional $\tilde E[h]$, which (as explained in section \ref{s:2})
receives a contribution from the fluid-solid interface:
\begin{eqnarray}\label{30} {\cal E}_{\rm bnry}\ =\ -
\oint_{\partial\cal D} \rmd l \, \gamma^\prime \, \tilde z \ =\ -
\gamma\cos\theta_0\, \int_{-\infty}^\infty \rmd{\sigma}\ \Bigl\{
\tilde z (1+\partial_\sigma \tilde y)\Bigr\vert_{\tau =0} - \tilde z
\Bigr\vert_{\tau =L_x/\sin\theta_0}\, \Bigr\} \ \ .
\end{eqnarray}
The second equality can be understood as follows: the unperturbed
planar surface meets the $x=0$, $x=L_x$ and $y=\pm L_y/2$ walls at
angles equal to $\theta_0$ , $\pi-\theta_0$ and $\pi/2$, respectively.
Young's equilibrium condition thus requires that, in the absence of
impurities: 
\begin{equation} \gamma^\prime = \cases{
&$\gamma\cos\theta_0$ \ \ \hskip 9mm {\rm for} \ \ $x=0$\ , \cr &$\ \
\ \ 0 $ \ \hskip 1.5cm {\rm for}\ \ $y=\pm L_y/2$\ , \cr &\hskip -2mm
$-\gamma\cos\theta_0$ \ \hskip 10mm {\rm for} \ \ $x=L_x$ \ .  }
\end{equation} 
Furthermore, along the first and the last wall the invariant length is
$\rmd l = \rmd y = (1 + \partial_\sigma \tilde y)\,
\rmd\sigma$. Dropping the quadratic term at $x=L_x$, since both
$\tilde y$ and $\tilde z$ must tend there to zero, gives the
advertised equation (\ref{30}).  Adding this to $\gamma \tilde {\cal
A}_{\rm min}$ leads to our final expression for the energy:
\begin{eqnarray}\label{enfina} \tilde E[h] = \ {\gamma\over 2}
\int_{-\infty}^\infty \rmd\sigma \, \Bigl[ \, i \tilde y_+ {\rmd
\tilde y_-\over \rmd\sigma} + i \tilde z_+ {\rmd \tilde z_-\over
\rmd\sigma} - \cos\theta_0\, \tilde z {\rmd \tilde y \over \rmd\sigma}
\, +\, c.c. \Bigr] \ .
\end{eqnarray}
Note that the linear terms cancel here separately on each wall, and
that all the contributions to the energy are ``quasi-local''.  Thus the
large-volume cutoff decouples, as expected, in the calculation of the
energy (but not of the separate contributions $\gamma \tilde {\cal
A}_{\rm min}$ and $ {\cal E}_{\rm bnry}$). The only restriction on the
cutoff is that it should not destabilize the unperturbed planar
surface.  We confirm these claims by a calculation in appendix B,
which includes as an extra control parameter the inclination angle of
the outer wall.  \vskip 0.5mm
 
For later use, we will also need the expression of the energy in terms
of the Fourier components of $\tilde y(\sigma)$ and $\tilde
z(\sigma)$.  Using eq.~(\ref{Laplace}), and doing some straightforward
algebra leads to 
\begin{equation}\label{NRJ} \tilde E[h] = 
\gamma \int_{0}^\infty  \frac{\rmd k}{2\pi}\,  k\, \Bigl( \, \vert \tilde
Y_k + i \cos\theta_{0} \tilde Z_k\vert^2 + \sin^2\theta_{0} \, \vert
\tilde Z_k\vert^2 \Bigr) \ .
\end{equation}
Note that the energy is quadratic in $\tilde y$ and $\tilde z$, where
the function $y(\sigma) = \sigma + \tilde y(\sigma)$ relates the
natural parameterization of the contact line, to the conformal
parameterization in terms of $\sigma$.  As was explained in the
previous section, the problem is non-linear because this change of
coordinate depends explicitly on the pinning profile.

%\vskip 3mm

%%%%%%%%%%%%%%%%%%%%%%%%%%%%%%%%%%%%%%%

\section{Perturbative expansion} 
\label{s:5} 

The pair of equations (\ref{bc1n}) and (\ref{constr}) cannot be
solved, in general, in closed form. However, if the contact line is
deformed only ``slightly'' (this will be made more precise later), then
$\tilde y$ and $\tilde z$ should both be small.  We may therefore
expand the right-hand side of eq.~(\ref{bc1n}) in a Taylor series,
\begin{eqnarray}\label{per1}
 {\tilde z}(\sigma)\, = \, \sum_{n=0}^\infty\, {{\rmd^n} h (\sigma
)\over \rmd\sigma^n} \, {{\tilde y (\sigma)}^{n}\over n!}\ ,
\end{eqnarray}
where both $\tilde y$ and the derivatives of $h$ are now evaluated at
the argument $\sigma$.  Furthermore, solving the quadratic equation
(\ref{constr}) for $\rmd \tilde y_+/\rmd\sigma$ and integrating gives:
\begin{eqnarray}\label{40}
 {\tilde y}_+(\sigma) = \int_{-\infty}^\sigma \rmd\sigma^\prime\,
\left\{ \left[ {1\over 4} - \Bigl({\rmd{\tilde z}_+ \over
\rmd\sigma^\prime}\Bigr)^2 - i \cos\theta_{0} {\rmd{\tilde z}_+ \over
\rmd\sigma^\prime} \right]^{1/2} -{1\over 2} \right\} \ \ .
\end{eqnarray}
Note that we have picked the solution of the quadratic equation that
vanishes for ${\tilde z}_+\to 0$, and we have also fixed arbitrarily
the irrelevant (complex) integration constant.  Since $\tilde z$ is
small, we may expand the integrand on the right-hand side to find
\begin{eqnarray} {\tilde y}_+(\sigma) &=& \sum_{n=1}^\infty
{2^{n-1}\over n!}\, [(-1)\cdot 1\cdot 3\cdot 5 \cdots (2n-3)]\times
\int_{-\infty}^\sigma \rmd\sigma^\prime\, \Bigl[ \Big({\rmd{\tilde
z}_+\over \rmd\sigma^\prime}\Big)^2 + i\cos\theta_{0} {\rmd{\tilde z}_+\over
\rmd\sigma^\prime}\Bigr]^n \nonumber \\
\label{pert1} &=& \int_{-\infty}^\sigma \rmd\sigma^\prime\, \Bigl[ -i
\cos\theta_0\, {\rmd{\tilde z}_+\over \rmd\sigma^\prime} \, -\,
\sin^2\theta_0\, \Big({\rmd{\tilde z}_+\over \rmd\sigma^\prime}\Big)^2
\ +\ \cdots \ \Bigr]
\end{eqnarray} 
Equations (\ref{per1}) and (\ref{pert1}) can now be solved iteratively
as follows: one starts with the lowest-order solution of the first
equation, $\tilde z(\sigma) = h(\sigma)$, and inserts it into the
second one to find ${\tilde y}_+ = - i\cos\theta_{0}\, h_+$.
Inserting the result in eq.~(\ref{per1}) gives $\tilde z$ at quadratic
order in $h$, and from eq.~(\ref{pert1}) we can obtain $\tilde y$ to
the same order.  Iterating the procedure gives, in principle, the
solution to any desired order in the pinning profile $h$.
 
\vskip 0.5mm In order to write the answer in a compact form, we
introduce the following notation.  If $f_\pm(\sigma)$ are the
positive- and negative-frequency parts of any real function
$f(\sigma)$, then 
\begin{equation} f \equiv f_+ + f_- \ \ \
{\rm and}\ \ \ \ i\, \widehat f := f_+ - f_-\ , 
\end{equation}
where the second equality defines the {\it dual} function $\widehat
f(\sigma)$.  Note that $f$ and $\widehat f$ are both real -- this
follows from the fact that $f_- = (f_+)^*$.  Now the first few orders
in the expansion of the solution read:
\begin{eqnarray}\label{c1} &&\tilde z  = h\, + \, \cos\theta_0\,
{\rmd h\over \rmd\sigma}\, \widehat h\, +\, {\cos^2\theta_0\over 2} \,
{\rmd^2 h\over\rmd\sigma^2}\, \widehat h^2 \, - \, {\rmd h\over
\rmd\sigma}\, \left[ \sin^2\theta_0 \, \int_{-\infty}^\sigma
\Big({\rmd h_+\over \rmd\sigma^\prime}\Big)^2 \rmd \sigma ' +
i\cos^2\theta_0\, \Bigl( {\rmd h\over\rmd\sigma}\widehat h\Bigr)_{\!+}
+ \, c.c. \right] \,
\, +O(h^4)\ , \qquad \\
\label{c2} &&\tilde y_+ = -i\cos\theta_0\, h_+\, - \, \sin^2\theta_0 \,
\int_{-\infty}^\sigma \Big({\rmd h_+\over \rmd\sigma^\prime}\Big)^2
\rmd \sigma ' - i\cos^2\theta_0\, \Bigl( {\rmd
h\over\rmd\sigma}\widehat h\Big)_{\!+} \, + \, O(h^3)\ ,
\end{eqnarray}
where we have stopped at one order lower in the expansion of $\tilde
y$ for a reason that will become apparent in a minute. It will be
useful to have also at hand the Fourier transforms of these
expressions. Noting that
\begin{equation} 
 f_+ g_+ + f_- g_- = {1\over 2} ( f g- {\widehat f}\, \widehat g ) \ \
\ \ \ {\rm and} \ \ \ \ \ i\, \widehat f_k = f_k \, {k\over \vert
k\vert}\ ,
\end{equation}
we find after some straightforward manipulations:
\begin{eqnarray}\label{res1}
i \tilde Y_k \, &=& \, h_k \, \cos\theta_0\, + \int h_{k_1}h_{k_2} \,
k_1 k_2 \, \Bigl[ \, {\sin^2\theta_0\over k}\, \Theta(k_1k_2) + \,
{\cos^2\theta_0\over \vert k_2\vert} \Bigr] \, +\, O(h^3)\ ,\\
 \tilde Z_k \, &=& \,
h_k  + \int h_{k_1} h_{k_2} \, k_1k_2 \, { \cos\theta_{0} \over
\vert k_2\vert} \, \
\nonumber \\
\label{res2} &&\hphantom{h_k}+ \int \, h_{k_1} h_{k_2} {h}_{k_3}\, k_1k_2k_3
\, \Bigl[ \, { \sin^2\theta_0 \over k_2+k_3}\, \Theta (k_2k_3) + {\cos^2\theta_0\, k_1 \over
2\vert k_2k_3\vert} + {\cos^2\theta_0\, (k_2+k_3)\over \vert k_3\vert
\vert k_2+k_3\vert} \Bigr] + \ O(h^4)\ .  
\end{eqnarray} 
Here the integrals run over all $k_j$, with normalization $\int \rmd
k_{j}/ (2\pi)$ and the condition that $\sum k_j = k$. The step
functions $\Theta(k_ik_j)$ force the two momenta to have the same
sign, and we have assumed that $k$ is positive.  Recall that $\tilde
Y_k$ enters into the expression (\ref{NRJ}) for the energy through the
combination
\begin{equation}\label{res3} i( \tilde Y_k + i \cos\theta_0 \, \tilde
Z_k) \, = \, { \sin^2\theta_{0}\over k} \, \int \, {h}_{k_1} {h}_{k_2}
\, k_1 k_2\, {\Theta(k_1k_2)}\, +\, O(h^3)\ .
\end{equation} 
Since this starts out quadratically in $h$, the cubic corrections
contribute to the energy at $O(h^5)$.  This explains why we have
truncated the expansion of $\tilde y$ at one order lower than the
expansion of $\tilde z$.

\vskip 0.5mm  
 
Inserting (\ref{res2}) and (\ref{res3}) in (\ref{NRJ}), and doing some
straightforward manipulations, leads to the following expression for
the energy of the deformed contact line at quartic order:
\begin{eqnarray} \tilde E[h] = E_2 + E_3 + E_4 \, +\, O(h^5)\ ,
\nonumber 
\end{eqnarray} 
where  \vskip - 4mm
\begin{eqnarray}\label{JDG} E_2 = {\gamma}\sin^2\theta_{0}\,
\int_{0}^\infty  \frac{\rmd k}{2\pi}\, k\, \vert h_k\vert^2\ \ ,
\end{eqnarray} 
\vskip - 2mm 
\begin{eqnarray}\label{e3} E_3 =
{\gamma}\cos\theta_0\, \sin^2\theta_{0}\, \int h_{k_1} h_{k_2} h_{k_3}
\, {\vert k_1\vert k_2k_3\over \vert k_3\vert}  \equiv  -
{\gamma}\cos\theta_0\, \sin^2\theta_{0}\, \int h_{k_1} h_{k_2} h_{k_3}\,
k_{1}k_{2}\, \Theta (k_{1}k_{2}) 
\end{eqnarray}
\vskip - 3mm
\begin{eqnarray}\label{inspect} E_4  = \frac{\gamma}2 \int
\, h_{k_1} h_{k_2} {h}_{k_3}h_{k_4}\, k_1k_2k_3 k_4 \, \Biggl[
\sin^4\theta_{0} \, \Bigl\{ { \Theta (k_1k_2)\Theta (k_3k_4) \over
\vert k_1+k_2 \vert } + {2k_1\over \vert k_1\vert}
{\Theta(k_3k_4)\over (k_3+k_4)} \Bigr\} \nonumber
\end{eqnarray} \vskip -4mm
\begin{eqnarray}\label{e4}
\ \ \ \ \ \ \ \ \ \ \ \ \ + \sin^2\theta_0 \cos^2\theta_0\, \Bigl\{
{k_1 k_4 \over \vert k_2k_3 k_4\vert} + { k_2^{\ 2}- k_1^{\ 2} \over
\vert k_1\vert \vert k_4\vert \vert k_1+k_2\vert } \Bigr\} \, \Biggr]
\ .
\end{eqnarray}
The integrals in (\ref{e3}) and (\ref{e4}) run over all $k_j$ with the
condition that $\sum k_j =0$.  As a check, note that for
$\theta_0=\pi/2$ the energy is invariant under reflection, $h\to -h$,
of the contact line.  Note also that the expressions multiplying
$\prod h_{k_j}$ inside the integrals are invariant under the
combination of complex conjugation and change of sign of all the
momenta, consistently with the fact that $\tilde E[h]$ should be real.
The expression for $E_3$ agrees with the one derived 
in \cite{GolestanianRaphael2001} by a different method. 
 
\vskip 0.9mm The Joanny-de Gennes linear theory
\cite{JoannyDeGennes1984} corresponds to the leading term of the above
expansion.  Comparing $E_2$ with the energy of an elastic rod, $E\sim
\int k^2 \vert h_k\vert^2$, one notes a softening of short-distance
modes, and corresponding hardening of long-distance modes, due to the
interactions mediated by the surface.  In real space, the JdG energy
can be written as (see the discussion in section \ref{s:3}):
\begin{equation}
E_2\, = \ {\gamma \over 4\pi} \, \sin^2\theta_0 \int \hskip -1.7mm \int \, d\sigma d\sigma'
 {[h(\sigma)-h(\sigma^\prime)]^2\over (\sigma -\sigma^\prime+ i\epsilon)^2}\ . 
\end{equation}
This quadratic, non-local functional has appeared in a variety of
other contexts, e.g.\ in simple models of quantum-mechanical
dissipation \cite{Caldeira,Callan}.  Note that $E_2$ is
invariant under SL(2,$\mathbb{R}$) transformations, i.e.\ under
conformal transformations that preserve the upper-half complex plane,
$\sigma \to \frac{a \sigma + b}{c \sigma + d}$, $h \to h$, with 
$a,b,c,d$ real and $a d - b c=1$. 
The full energy is not only translationally-invariant, but it also
transforms covariantly under rescalings of the physical space:
\begin{equation}\label{rescale}
\tilde E [h^{(\lambda)}]\ =\ \lambda^2\, \tilde E[h] \ \ \ \ \ \ \
{\rm if}\ \ \ \ \ \ \ \ h^{(\lambda)}(y) \equiv \lambda\, h
(\lambda^{-1} y)\ \ .
\end{equation}
This implies that the perturbative expansion is really an expansion in
derivatives, as should be expected from the fact that the classical
problem has no intrinsic length scale. We will return to this point
later on.

\vskip 0.5mm

It will be useful, for comparison with the following section, to
rewrite the quartic contributions to the energy differently. First, we
note that the two terms multiplying $\sin^4\theta_0$ are equal up to a
factor of $-2$. This follows from the following chain of replacements,
which are allowed upon symmetrization of the integrand:
\begin{eqnarray}
 {2k_1\over \vert k_1\vert} {\Theta(k_3k_4)\over (k_3+k_4)} \ \ \
\longrightarrow \ \ \ -(s_1 + s_2)\, {\Theta(k_3k_4)\over (k_1+k_2)} \
\ \ \longrightarrow \ \ \ -(1+s_1s_2) \, { \Theta(k_3k_4)\over \vert
k_1+k_2\vert} \nonumber
\end{eqnarray}
Here $s_j = k_j/\vert k_j\vert$ is the sign of the momentum $k_j$, and
in the second step we have used the fact that the sign of $(k_1+k_2)$
is the same as the sign of either $k_1$ or $k_2$, since the expression
is multiplied by $(1+s_1s_2) = 2\Theta(k_1k_2)$.  Likewise, one can
justify the following replacement:
\begin{eqnarray}
 {k_1\over \vert k_1\vert} {\Theta(k_3k_4)\over (k_3+k_4)} \ \ \
\longrightarrow \ \ \ { k_1k_4 \over \vert k_1\vert \vert k_4\vert
\vert k_1+k_2\vert } \ , \ \ \ {\rm and}\ \ \ { k_2^{\ 2}- k_1^{\ 2}
\over \vert k_1\vert \vert k_4\vert \vert k_1+k_2\vert } \ \ \
\longrightarrow \ \ \ { (k_2^{\ 2} + k_3^{\ 2}) - (k_1^{\ 2}+ k_4^{\
2})\over 2 \vert k_1\vert \vert k_4\vert \vert k_1+k_2\vert } \ \ .
\nonumber
\end{eqnarray} 
Putting all these facts together, using that $\sum_j k_j =0$
and doing some straightforward rearrangements leads to the following
alternative expression for the quartic energy:
\begin{eqnarray}\label{e4n} E_4  = \frac{\gamma}2 \sin^2\theta_{0} \int
\, \prod_{j=1}^4 (k_j h_{k_j} ) \, \Biggl[ - { \Theta (k_1k_2)\Theta
(k_3k_4) \over \vert k_1+k_2 \vert } + \cos^2\theta_0\, \Bigl\{ {k_1
k_4 \over \vert k_2k_3 k_4\vert} - { k_2k_3 \over \vert k_1\vert \vert
k_4\vert \vert k_1+k_2\vert } \Bigr\} \, \Biggr] \ .
\end{eqnarray}
This somewhat more economical expression will be easier to  compare with the
diagrammatic expansion, to which we will now turn our attention. Note that the expression for
$E_4$ in the particular case $\theta_0=\pi/2$ was also found in \cite{LeDoussalWieseRaphaelGolestanian2004}
using the perturbative solution of the non-linear equation (not using conformal coordinates). It is possible, though cumbersome, to extend the method to arbitrary $\theta_0$ \cite{LeDoussalGolestanian}.

%%%%%%%%%%%%%%%%%%%%%%%%%%%%%%%%%%%%%%%%%

\section{Diagrammatic method} 
\label{s:6}
The perturbative expansion of the energy can be organized efficiently
by using a Lagrange-multiplier field to impose the pinning constraint
of the contact line.  One starts with the following variational
principle for the area:
\begin{eqnarray}\label{vrt}
 {\cal A}_{\rm min} \, = \, {\rm extr}\, {\cal A}(\alpha, \vec r)\ , \
\ \ {\rm with} \ \ \ \ {\cal A}(\alpha, \vec r) = \int\hskip -2mm
\int_{{\cal D}} \rmd^2\sigma \sqrt{{\rm det} g}\, - \, \int_{\partial
{\cal D}} \rmd s \, { \alpha(s)} \Bigl[z (s) - h\bigl( y(s)\bigr)
\Bigr]\ .
\end{eqnarray}
Here $s$ parameterizes the boundary of the domain ${\cal D}$, and
$\alpha$ is a Lagrange-multiplier field that transforms under
reparametrizations such that $\alpha(s)\, \rmd s$ remains
unchanged. Since ${\cal A}(\alpha, \vec r)$ is
reparametrization-invariant, we are free to choose the conformal gauge
and to set $x= \sin\theta_0\,\tau$\ \ as before. Thus ${\cal D}$ is
the upper-half plane $\tau \geq 0$, and we may choose $s= \sigma$ for
the boundary parameter. We also define $y = \sigma +\tilde y$ and $z =
-\cos\theta_0\, \tau +\tilde z$ , and we subtract from ${\cal A}$ the
area of the flat fluid surface. This gives $\tilde {\cal A}_{\rm min}
\, = \, {\rm extr}\, \tilde {\cal A}$, where
\begin{eqnarray}\label{vart}
  \tilde {\cal A}(\alpha, \tilde y, \tilde z) = {1\over 2}\,
\int\hskip -2mm \int_{{\tau\geq 0}} (\partial_a \tilde y\,
\partial^a\tilde y + \partial_a \tilde z \, \partial^a\tilde z) \, -
\, \int_{{\tau = 0}} \, \Bigl[ {\alpha} \, \bigl(\tilde z - h(\sigma
+\tilde y) \bigr) \, -\, \cos\theta_0 \, \tilde z \Bigr] \ .
\end{eqnarray}
The last term in the above expression comes from the cross term
$\partial_a z_0\,\partial^a\tilde z = - \cos\theta_0 \,\partial_\tau
\tilde z$\ \ in the area difference.  This is a total derivative,
which is why it only contributes a boundary term.  Note that, in the
light of our discussion in section \ref{s:4}, all contributions from
the boundaries at infinity have been dropped. This is legitimate since we
are ultimately interested in the energy (\ref{enfina}) rather than in
the area of the fluid surface. Alternatively, one can view
$\tilde {\cal A}(\alpha, \vec r)$ as an action and consider the path integral 
over the fields $\vec r$ and $\alpha$ \footnote{For calculational convenience the choice of convention here - see the propagators in e.g. Eq. (\ref{a3}) - corresponds to a weight
$\exp(\tilde {\cal A}(\alpha, \vec r))$. This choice is immaterial since we are 
doing only a tree level calculation. One can equally well use 
the more physical choice $\exp(- \tilde {\cal A}(\alpha, \vec r))$ with
$\alpha \to i \alpha$ and positive signs in all propagators, with identical final results at tree level.}.
Since we are doing only a tree level calculation there is no need to worry about
Fadeev-Popov ghosts, which would be important for the study of
thermal or quantum fluctuations.  Fluctuating surfaces \cite{david}
are beyond the scope of the present study.  \vskip 0.5mm

It looks, at first sight, rather odd that in the above formulation the
conformal-gauge conditions are not explicitly imposed.  The extrema of
$\tilde{\cal A}(\alpha,\tilde y, \tilde z)$ should therefore obey
these conditions automatically.  To see why, note that the variation
of (\ref{vart}) leads to the boundary equations:
\begin{eqnarray}\label{vartb}
\partial_\tau\tilde y = {\alpha}(\sigma)\, h^\prime(\sigma +\tilde y)\
\ \ \ {\rm and}\ \ \ \partial_\tau\tilde z = - {\alpha(\sigma)} +
\cos\theta_0\ \ \ {\rm at}\ \ \ \tau =0\ .
\end{eqnarray}
From the above boundary equations, and from the pinning
constraint $z=h(y)$, we deduce:
\begin{eqnarray}\label{vartb1}
\partial_\tau \vec r \, =\, \bigl(\sin\theta_0\, ,\, {\alpha
(\sigma)}\, h^\prime(y)\, , \, -{\alpha (\sigma)} \bigr) \ \ \ {\rm
and}\ \ \ \ \ \partial_\sigma \vec r \, =\, \bigl(0\, , \,
\partial_\sigma y\, ,\, h^\prime(y)\, \partial_\sigma y\bigr)\ .
\end{eqnarray}
Thus, on the boundary, the condition $\partial_\sigma\vec r\cdot
\partial_\tau\vec r = 0$ holds.  This implies that the function
$\partial_w \vec r \cdot \partial_w\vec r$, which is analytic in the
upper half plane and vanishes at infinity, has zero imaginary part on
the real axis. From the Cauchy-Poisson integral formula \cite{compl}
we conclude that it vanishes everywhere, so that the conformal gauge
conditions (\ref{conf4n}) are indeed satisfied.

\vskip 0.5mm In order to develop simple diagrammatic rules, we first
solve the harmonic equation for the ``bulk'' fields keeping their
restrictions to the boundary, $\tilde y(\sigma) := \tilde
y(\sigma,0)$ and $\tilde z(\sigma) := \tilde z(\sigma,0)$, fixed. As
has been already discussed, this leads to the replacement
\begin{eqnarray}\label{repla}
  {1\over 2}\, \int\hskip -2mm \int_{\tau\geq 0} \partial_a \tilde y\,
\partial^a\tilde y \ \ \longrightarrow \ \ {i\over 2}\, \int_{\tau=0}
(\tilde y_+ {\rmd \tilde y_-\over\rmd\sigma} -\tilde y_- {\rmd \tilde
y_+\over\rmd\sigma} )\, = \, \frac{1}{2} \int_k \vert k\vert \, \tilde y_k\,
\tilde y_{-k} \ \ ,
\end{eqnarray}
and likewise for the field $\tilde z$. Next, we solve the linear
equations for $\tilde z(\sigma)$, thus eliminating it entirely from
the expression (\ref{vart}). The new variational functional, expressed
in terms of Fourier components, reads 
\begin{eqnarray}\label{fnct}
 \tilde {\cal A}(\alpha, \tilde y)\, = \, {1\over 2}
\int_k \vert k\vert \, \tilde y_k\, \tilde y_{-k} \, -\, {1\over 2}
\int_k {1\over \vert k\vert} \, \delta\alpha_k \, \delta\alpha_{-k} \,
+ \, \int_k \alpha_{-k} H_k \ ,
\end{eqnarray} 
where\ $\alpha_k = \cos\theta_0\, 2\pi \delta(k) + \delta\alpha_k$\ , and
$H_k$ is the Fourier transform of $H(\sigma) = h(\sigma+\tilde
y(\sigma))$. This result also follows if one uses the path integral formulation and integrates
over the fields $\tilde y$ and $\tilde z$ 
in the bulk. More explicitly
\begin{equation}
H_k = h_k + \int ik_1\, h_{k_1}\, \tilde y_{k_2} + {1\over 2}\int
(ik_1)^2 h_{k_1}\, \tilde y_{k_2} \, \tilde y_{k_3} + \cdots \ \ ,
\end{equation}
where the integrals run over $\sum k_j = k$.  The extremum of the
functional (\ref{fnct}) can be computed by summing tree-level diagrams
of a 1-dimensional field theory. The 1-point function and
propagators read: 
\begin{eqnarray}\label{a3} 
\diagram{dcostheta}\ &:= &\ \left<\, \alpha_k \, \right>\ =\ \cos
 \theta_0\, 2\pi \delta(k) \nonumber \\
  \diagram{line}\ &:=&\ \left<\,  \delta\alpha_k\,  \delta\alpha_{-k}\,
 \right>\ =\ |k|  \\
 \diagram{wiggly} \ &:=&\ \left< \tilde y_k\, \tilde y_{-k}\right>\ =\
-\frac{1}{|k|}\nonumber
\end{eqnarray} 
while the first few vertices, deriving from the last term of
(\ref{fnct}), are as follows:
\begin{equation}\label{a4}
\diagram{vertex1}:= \ h_k \alpha_{-k} \ , \qquad \diagram{vertex2} :=
\ ik_1\, h_{k_1}\, \tilde y_{k_2}\, \alpha_{-k_1-k_2} \ , \qquad
\diagram{vertex3}:= \ -\frac{(k_1)^2}{2} \, h_{k_1}\, \tilde y_{k_2}\,
\tilde y_{k_3}\, \alpha_{-k_1-k_2-k_3} \ .
\end{equation}
Note that all of these vertices are proportional to the amplitude of
the pinning profile.  Furthermore wiggly lines, corresponding to the
field $\tilde y$, can only terminate on another vertex in a vacuum
tree diagram.  Thus only a finite number of vertices  contributes
to a given order in the expansion in $h$.  Solid lines corresponding
to the Lagrange-multiplier field $\alpha$ may end
at the tadpole  $\left<\alpha_k \right> = \cos\theta_0\, 2\pi \delta(k)$,
which carries no extra power of $h$.  Note also that at the vertices
momentum is  injected by $h_k$, which has to be taken into account for
momentum conservation.

\vskip 0.5mm Using the above diagrammatic rules, one can compute any
desired order in the expansion of $\tilde E[h]$.  This is obtained by
multiplying the extremum of (\ref{fnct}) with $\gamma$, and then
subtracting the linear contribution of the wall, ${\cal E}_{\rm bnry}
= \gamma^\prime\int h = \gamma\cos\theta_0\, h_0$ (see section
\ref{s:4}).  This contribution cancels precisely the tadpole diagram
\begin{equation}\label{a5} 
\diagram{ord1}\ =\ \cos \theta_0 \, h_0 \ \, 
\end{equation} 
in agreement with the fact that the unperturbed,
planar fluid surface should be stable.  Denoting by $E_n$ the
$n$th-order term in the perturbative expansion of $\tilde E[h]$ one
finds:
 
\bigskip 
\noindent {\underline {Order 2}:}
\begin{equation}\label{a6}
 \diagram{dwig0} \ =\ \frac{1}{2} \int_{k} |k|\,  h_{k}\,  h_{-k} 
\ , \qquad 
 \diagram{ord2a}\ = \ - \frac{1}{2}\cos^2\theta_0  
 \int_{k}\frac{k^{2}}{|k|} \, h_{k}\,  h_{-k} \ , 
\end{equation}
so that
\begin{equation}\label{a7} E_2 \ =\
\gamma\, \Biggl( \diagram{dwig0}+ \diagram{ord2a} \Biggr)\ = \
\frac\gamma 2 \sin^2\theta_0 \int_{k} |k|\, h_{k} h_{-k} \ \ ,
\end{equation}
\vskip 1.5mm \noindent which is precisely the Joanny-de Gennes
quadratic energy.  

\bigskip 
\noindent 
{\underline {Order 3}:} 
\begin{equation}\label{4.19}
\diagram{ord3a}\ =\ \cos\theta_0 \int  \frac{|k_{1}|k_{2}
k_{3}}{ |k_{3}|} \,  h_{k_1} h_{k_2} h_{k_3}
 \end{equation}
\begin{equation}\label{4.20}
\diagram{ord3b}\ =\ \frac{1}{2}  \cos^3\theta_0 \int 
\frac{k_{1}k_{2}^{\ 2} k_{3}}{|k_{1}| |k_{3}|}\,   h_{k_1} h_{k_2}  h_{k_3}
\end{equation}
\vskip 1.5mm \noindent These two contributions are of the same
form. To see why, one must symmetrize the integrands over all
permutations of [123], and then use the identities that follow from
momentum conservation:
\begin{eqnarray}
{1\over 2}\,  s_1 s_3\,  k_2^{\ 2} + {\rm perms}\ =\ 
s_1 s_3\,  (k_1^{\ 2} + k_1k_3) + {\rm perms}\ =\ 
-s_1s_3\,  k_1 k_2 + {\rm perms} \ ,   \nonumber
\end{eqnarray}
where $s_j = k_j/\vert k_j\vert $ is the sign of $k_j$. Thus, the sum
of the two diagrams gives 
\begin{equation}\label{a8} E_3\ =\
\gamma\, \Biggl( \diagram{ord3a}+\diagram{ord3b} \Biggr) \ = \
\gamma\, {\cos \theta_0}\, \sin^2 \theta_0 \int \frac{|k_{1}|k_{2}
k_{3}}{ |k_{3}|} \, h_{k_1} h_{k_2} h_{k_3}\ ,
\end{equation} 
\vskip 1.5mm \noindent which agrees with the calculation (\ref{e3}) of
the previous section.
 
%%%%%%%%%%%%%%%%%%%%%%

\bigskip \noindent 
{\underline {Order 4}: }  
%\begin{equation}\label{a14}
 %{\cal E}^{(4)} = \diagram{dwig1}
 %+\diagram{ord4a}+\diagram{ord4b}+\diagram{ord4c}+\diagram{ord4d}
%\end{equation}

\begin{eqnarray}\label{maeh}
  \diagram{dwig1} \ = \ \frac{1}{2}
\int  \frac{|k_1| k_2 k_3 |k_4|}{|k_1+k_2|} h_{k_1}
h_{k_2} h_{k_3} h_{k_4}  
\end{eqnarray}
\begin{equation}\label{a12}
 \diagram{ord4c}\ =\ \frac{1}{2}  \cos^2\theta_0
\int\,\frac{k_{1}k_{2}^{2}k_{3}|k_{4}|}{|k_{1}\vert\vert k_{3}|} \, 
h_{k_1} h_{k_2} h_{k_3} h_{k_4}
\end{equation}
\begin{equation}\label{a11}
 \diagram{ord4b}\ =\ \cos^2\theta_0  \int \frac{k_{1}
k_{2}^{2}  k_{3}  |k_{4}|  }{|k_{1}  |\, |  k_{1}+k_{2}|  }\,  h_{k_1}
h_{k_2} h_{k_3} h_{k_4}
\end{equation}
\begin{equation}\label{a13}
 \diagram{ord4d}\ =\ \frac{1}{2}  \cos^2\theta_0 \int\,
\frac{k_{1}k_{2}k_{3}k_{4}|k_1+k_{2}|}{|k_{1}\vert\vert k_{4}|}\, 
h_{k_1} h_{k_2} h_{k_3} h_{k_4} 
\end{equation}
\begin{equation}\label{a9}
 \diagram{ord4a}\ =\ \frac{1}{6} \cos^4\theta_0 \int
 {k_1k_2k_3k_{4}^{3}\over \vert k_1\vert \vert k_2\vert \vert k_3\vert }
 \, h_{k_1} h_{k_2} h_{k_3} h_{k_4}
\end{equation}
\begin{equation}\label{a10}
 \diagram{ord4f}\ =\ \frac{1}{2} \cos^4\theta_0 \int 
\frac{k_{1}k_{2}^{2}k_{3}^{2}k_{4}}{ |k_{1}||k_{1}+k_{2}||k_{4}|}\,
h_{k_1} h_{k_2} h_{k_3} h_{k_4}
\end{equation}
\vskip 1.8mm \noindent Note that the power of $\cos\theta_0$
corresponds to the number of ``hooks'' of a given
diagram. 
 For ${\rm cos}\theta_0 = 0$ only the first of these diagrams
contributes.  Using the replacements
\begin{eqnarray}
{s_1s_4\over \vert k_1+k_2\vert } \ \ \longrightarrow \ \
{(s_1+s_2)(s_3+s_4) \over 4\vert k_1+k_2\vert} \ =\ - {\Theta(k_1 k_2)
\Theta(k_3 k_4) \over \vert k_1+k_2\vert}\ , \nonumber
\end{eqnarray}
one can check that this diagram agrees with our previous result
(\ref{e4n}).  To show that the expression (\ref{e4n}) also agrees with
$E_{4}=\gamma \left[
(\ref{maeh})+ (\ref{a12})+ (\ref{a11})+(\ref{a13})+ (\ref{a9})+
(\ref{a10})\right]$   for arbitary contact angle $\theta_0$ we 
proceed as follows:  first 
 diagrams (\ref{a12}) and (\ref{a9}) can be
combined to reproduce the second term in (\ref{e4n}).  To see why, one
must replace\ \ ${1\over 3} k_4^{\ 2} = - {1\over 3} k_4 (k_1+k_2+k_3)
\ \to\ - k_1 k_4$ in the integrand of the diagram (\ref{a9}).  
Secondly, one can show that  for ${\rm cos}\theta_0 =1$ 
the sum of the remaining diagrams,  $(\ref{maeh})+ (\ref{a11}) +
(\ref{a13}) + (\ref{a10}) $,  is exactly zero.   Indeed, writing 
the integrands of these diagrams  in the order of their appearance  we find:
\begin{eqnarray}\label{c3}
&&\!\!\!\frac{k_{1}k_{2}k_{3}k_{4}}{|k_{1}|\, |k_{4}|\,|k_{1}+k_{2}|}
\left[k_{1}k_{4} + 2 k_{2}k_{4}+ (k_{1}+k_{2})^{2}+  k_{2}k_{3} \right]
= \frac{k_{1}k_{2}k_{3}k_{4}}{|k_{1}|\, |k_{4}|\,|k_{1}+k_{2}|}
\left[k_{1}k_{4} +  2 k_{2}k_{4}- (k_{1}+k_{2})
(k_{3}+k_{4}) + k_2k_3 \right] \nonumber \\
&& \qquad = \frac{k_{1}k_{2}k_{3}k_{4}}{|k_{1}|\,
|k_{4}|\,|k_{1}+k_{2}|} \left(k_{2}k_{4}-k_{1}k_{3} \right) =\frac{k_{1}k_{2}k_{3}k_{4}}{|k_{1}|\,
|k_{4}|\,|k_{3}+k_{4}|} \left(k_{2}k_{4}-k_{1}k_{3} \right) \ ,
\end{eqnarray}
where in the first and the third equality we have used the
conservation of momentum.  Since the last two expressions are equal
they can be replaced by their average.  The result is antisymmetric
under the exchange of 1 with 4 and 2 with 3, so after multiplication
with $h_{k_1}h_{k_2}h_{k_3}h_{k_4}$ it gives zero as claimed. We are
thus free to subtract this vanishing expression times $\frac{1}{2}
\cos^{2} \theta_{0} $ from the sum of all diagrams. This removes the
contributions (\ref{a11}) and (\ref{a13}), and changes the
coefficients of (\ref{maeh}) and (\ref{a10}) to those of the
corresponding terms in (\ref{e4n}). This completes the proof that
equation (\ref{e4n}) agrees with the diagrammatic calculation of the
energy.  \vskip 0.7mm

The diagrammatic expansion can be extended to higher orders.  As an
illustration, let us consider the case of a perpendicular contact
angle, in which case the tadpole vanishes.  The symmetry under $h\to
-h$ guarantees that only even powers appear in the expansion of $E[h]$.
The sixth and eighth order terms are given by:
\begin{eqnarray}\label{e6n}
E_6 \Bigl\vert_{\theta_0=\pi/2}  \ = \ \gamma\ 
\diagram{dwig2} \ =\ \gamma\ \int  \frac{|k_1| k_2 k_3^2 |k_4| k_5
|k_6|}{|k_1+k_2| |k_5+k_6|}  \ \prod_{j=1}^6 h_{k_j}
\end{eqnarray}
\begin{eqnarray}
E_8\Bigl\vert_{\theta_0=\pi/2} = \gamma\ \Bigl(\ \diagram{dwig3} +
\diagram{dwig4}\ \Bigr) \nonumber
\end{eqnarray}
\begin{eqnarray}\label{e8n}
 {\rm where}\ \ \ \ \ \ \ \ \ \ \diagram{dwig3} = \int \frac{|k_1|\,
k_2 k_3^2\, |k_4|\, k_5^2\, |k_6|\, k_7\, |k_8|}{|k_1+k_2|
|k_1+k_2+k_3+k_4| |k_7+k_8|} \ \prod_{j=1}^8 h_{k_j} \end{eqnarray}
\begin{eqnarray}\label{endd}
{\rm and}\ \ \ \ \ \ \ \ \diagram{dwig4} = \int \frac{|k_1|\, k_2\,
|k_3|\, k_4\, |k_5| \, k_6\, |k_7|\, k_8^3\, }{|k_1+k_2| |k_3+k_4|
|k_5+k_6|} \ \prod_{j=1}^8 h_{k_j}
\end{eqnarray}
We will comment further on these results in the final section.

\section{Numerical algorithm}\label{s:C}
As was shown in section \ref{s:4}, the problem of determining the
deformed fluid surface with a pinned contact line on a planar wall
reduces to that of solving the pair of equations for the real
functions $\tilde y(\sigma)$ and $\tilde z(\sigma)$:
\begin{equation}\label{bc1napp} \tilde z(\sigma) = h \Bigl(\sigma +
\tilde y(\sigma)\Bigr)\ \ \ \ \ \ \ \ {\rm and}\ \ \ \ \ \ \ \ {\rmd
\tilde y_+\over \rmd\sigma} + i \cos\theta_0 \, {\rmd \tilde z_+\over
\rmd\sigma} = - \left({\rmd \tilde y_+\over \rmd\sigma}\right)^{\!2} -
\left({\rmd \tilde z_+\over \rmd\sigma}\right)^{\!2}\ .
\end{equation}
We recall that $f_\pm$ are the projections onto positive (negative)
momentum Fourier components of the function $f$. Continuing $f_+$ as
an analytic function, $ F $, in the upper-half $w=(\sigma+i\tau)/2$
plane, determines the unique harmonic extension of the function, $f =
2\, {\rm Re} F$. The conformally-parameterized minimal surface is
\begin{equation}
(x , y , z ) = ( \sin\theta_0\, \tau, \sigma + \tilde y ,
-\cos\theta_0\, \tau + \tilde z\, )\ ,
\end{equation}
and it has a total energy given by eqs.~(\ref{enfina}) or (\ref{NRJ}).
\vskip 0.5mm The equations (\ref{bc1napp}) can be solved by iteration,
starting with the initial configuration
\begin{eqnarray} {\tilde
y}^{(0)}(\sigma) = 0 \ \ , \ \ \ \ {\tilde z}^{(0)} (\sigma) = h
(\sigma)\ .  
\end{eqnarray} 
Let ${\tilde y}^{(n)}$ and ${\tilde z}^{(n)}$ be the solution of the
equations after $n$ steps.  We extract ${\tilde z}_+^{(n)}$ by doing a
double Fourier transform.  Plugging the result in eqs.
(\ref{bc1napp}) then gives the improved values of the unknown
functions:
\begin{eqnarray}\label{40app} {\tilde y}_+^{(n+1)} =
\int_{-\infty}^\sigma d\sigma^\prime\, \left\{ \left[ {1\over 4} -
\Big({d{\tilde z}_+^{(n)}\over d\sigma^\prime}\Big)^2 - i \cos\theta_{0}
{d{\tilde z}_+^{(n)} \over d\sigma^\prime} \right]^{1/2} -{1\over 2}
\right\} \ \ , \ \ \ \ {\tilde z}^{(n+1)} = h \left(\sigma + {\tilde
y}^{(n+1)} \right)\ .  
\end{eqnarray} 
Using (\ref{NRJ}) yields an approximation $E_{n}$ to the true energy
$E_{\infty}$.  We have used this iterative algorithm for $h(y) =
\varepsilon\, f(y)$ with $f(y)$ various trial pinning profiles.  We
found that it converges rapidly to the perturbative result for small
$\varepsilon$, and that it breaks down at some critical $\varepsilon$
where the function $y(\sigma)$ stops being monotonic.  We believe this
signals a coordinate, rather than a real geometric singularity, as is
observed in section \ref{s:Weierstrass}. If so, it would be very
interesting to develop alternative algorithms that could circumvent
this problem.
 
On figure \ref{f:algo1} we show the convergence of the algorithm at
$\theta =\pi /2$ for a profile $h (y)$ given on figure \ref{f:algo2},
together with the corresponding functions $y (\tau)$ and $z
(\tau)$. One sees on figure \ref{f:algo2} already the emergence of a
linear cusp at the tip of $z (\tau=1/2)$, which signals for larger
$\epsilon$ the break-down of the algorithm.

\begin{figure}
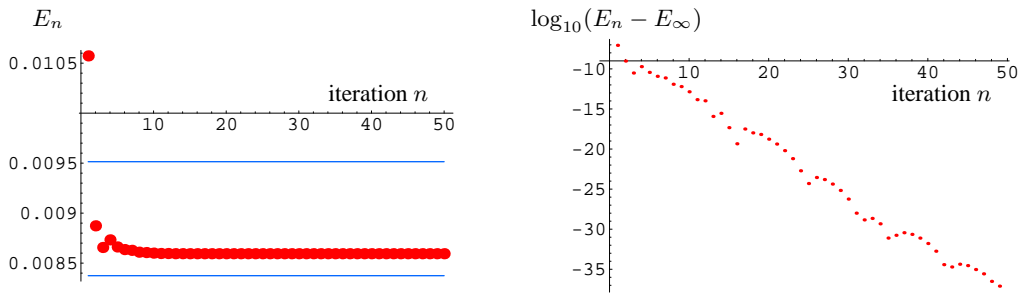

$\rule{0mm}{3.8cm}^{\displaystyle
E_{n}}\hspace{-0.8cm}\fig{6cm}{convergelin}\hspace{-1.7cm}\rule{0mm}{2.8cm}^{\mbox{iteration~$n$}}$
\qquad\qquad $\rule{0mm}{3.8cm}^{\displaystyle \log_{10} (
E_{n}-E_{\infty
})}\hspace{-1.8cm}\fig{6cm}{convergelog}\hspace{-1.7cm}\rule{0mm}{2.8cm}^{\mbox{iteration~$n$}}$
\caption{Convergence of the energy $E_{n}$ for a Gaussian with almost
maximal amplitude, as function of iteration $n$. Also plotted are the
perturbative results $E_{2}=0.00951444$, and
$E_{2}+E_{4}=0.00837429$. The second plot shows convergence on a
$\log_{10}$ scale. Convergence improves considerably for smaller
amplitude of $h$.} \label{f:algo1}
\end{figure}
\end{widetext}

\begin{figure}
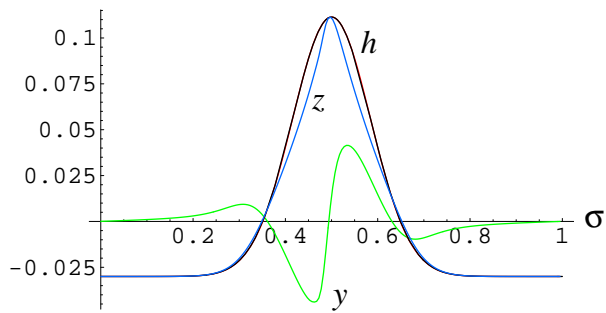

\fig{8cm}{NewtonGaussXfig} \caption{A periodically repeated Gaussian
for $h (y)$, with no 0-mode: $\int_{0}^{1}\rmd y\, h (y) =0$. The
corresponding functions $y (\sigma)$ and $z (\sigma)$ on the boundary
are also given. One remarks that $z$ almost has a cusp-like
singularity at $\sigma =1/2$. Further increasing the amplitude of $h$,
$z$ will develop this cusp, which signals the breakdown of our
parameterization.}  \label{f:algo2}
\end{figure}

%%%%%%%%%%%%%%%%%%%%%%%%%%%%%%%%%%%%%%%

%\vskip 3mm 

\section{Interaction between contact lines} 
\label{s:7} 

As another application of the general approach, we will now calculate
the interaction between the two contact lines of a liquid surface
bounded by parallel walls.  For an analogous calculation in open
string theory see reference \cite{bg}.  Suppose that wall 1 is located
at $x=0$, wall 2 at $x = L$, and let $\gamma^\prime_1 =
-\gamma^\prime_2 = \gamma \cos\theta_0$.  In the absence of impurities
the equilibrium configuration is thus an inclined planar surface
making a contact angle $\theta_0$ (respectively $\pi -\theta_0$) with
the first (second) wall.  We use conformal coordinates and set $x =
\sin\theta_0\, \tau$, so that the parameter domain is the infinite
strip $0\leq\tau\leq L/\sin\theta_0 \equiv \tau_0$.  Repeating the
same steps as in the previous section leads to the following
variational functional for the minimal area:
\begin{eqnarray}\label{vart1} \tilde {\cal A}(\alpha_K, \tilde y,
\tilde z)\ = \ {1\over 2}\, \int\hskip -2mm \int_{{0\leq \tau\leq
\tau_0}} (\partial_a \tilde y\, \partial^a\tilde y + \partial_a \tilde
z \, \partial^a\tilde z) \, - \, \int_{{\tau = 0}} \, \Bigl[
{\alpha}_1 \, \bigl(\tilde z - h_1(\sigma +\tilde y) \bigr) \, -\,
\cos\theta_0 \, \tilde z \Bigr] \nonumber 
\end{eqnarray} 
\vskip -4mm
\begin{eqnarray} \ \ \ \ \ \ \ \ \ \ \ \ \ \ \ \ \ \, - \, \int_{{\tau
= \tau_0}} \, \Bigl[ {\alpha}_2 \, \bigl(\tilde z - h_2(\sigma +\tilde
y) \bigr) \, + \, \cos\theta_0 \, \tilde z \Bigr] \ .
\end{eqnarray}
Here $h_J(y)$ (for $J=1,2$) are the deformations of the two contact
lines away from their equilibrium configuration, and $\alpha_J$ are
the corresponding Lagrange-multiplier fields. The minimal area
difference is $\tilde {\cal A}_{\rm min} \, = \, {\rm extr}\, \tilde
{\cal A}$ , where one must extremize $\tilde {\cal A}$ over the bulk
fields $\tilde y(\sigma ,\tau)$ and $\tilde z(\sigma , \tau)$ and the
boundary fields $\alpha_J(\sigma)$.

\vskip 0.5mm

First we solve the harmonic equations for $\tilde y$ and $\tilde z$,
keeping their values on the boundary fixed.  Let, for example,
$ \tilde y(\sigma, 0) = \tilde y_1(\sigma)$ and $ \tilde
y(\sigma,\tau_0) = \tilde y_2(\sigma)$. Eliminating the field in the
interior gives
\begin{eqnarray}\label{vartn}
{1\over 2} \int \hskip -2mm \int_{0\leq\tau\leq\tau_0} \partial_a
\tilde y\, \partial^a \tilde y\ \ \longrightarrow\ \ {1\over 4\pi}\,
\sum_{J, J^\prime} \, \int _\sigma \hskip -1 mm \int_{\sigma^\prime }
\, {\rmd \tilde y_{J}\over \rmd\sigma} \, G_{J J^\prime}(\sigma -
\sigma^\prime ) \, {\rmd \tilde y_{ J^\prime}\over \rmd\sigma^\prime}\
, \end{eqnarray} where
\begin{eqnarray}\label{cas}
 G_{J J^\prime}(\sigma -  \sigma^\prime)
 \ =\ \cases{\ 
 - {\rm log}\,   {\rm sinh}^2[{\pi  \over 2\tau_0} (\sigma-\sigma^\prime
 )]
\ \ & {\rm if}\ \ $J=J^\prime$\ , 
 \cr  & \,   \cr\ 
 - {\rm log}\,   {\rm cosh}^2[{\pi  \over 2\tau_0} (\sigma-\sigma^\prime
 )] 
\ \ & {\rm if}\ \ $J\not= J^\prime$\ .}
\end{eqnarray}
One way of establishing this formula, is to start from the analogous
expression for the unit disk, eq.~(\ref{minan}), and then apply the
conformal transformation that maps the unit disk onto the infinite
strip:
\begin{equation}\label{ctrans}
v \equiv \sigma + i\tau = \ {\tau_0\over\pi}\, {\rm log}\left( i\, {1-
w\over 1+w}\right)\ \ \ \Longleftrightarrow\ \ w \equiv \rho e^{i\phi}
= \, {\rm tanh} \left( {i\pi\over 4} - {\pi v\over 2\tau_0}\right)\ .
\end{equation}
Notice that the two unit-radius semi-circles, $\rho =1$ and $\phi\in
[0, \pi]$ or $\phi\in [\pi, 2\pi]$, are indeed mapped onto the two
boundaries of the strip, ${\rm Im} v =0$ or ${\rm Im} v = \tau_0$.  On
these boundaries
\begin{equation}\label{c4}
  {\rm log}\, {\rm sin}^2 \left({\phi-\phi^\prime\over 2} \right) \ =
\ {\rm log}\, {\rm sinh}^2\left[{\pi \over 2\tau_0} (v- v^\prime
)\right] - {\rm log}\, \cosh \left({\pi v\over\tau_0}\right) - {\rm
log}\, \cosh \left({\pi v^\prime \over\tau_0}\right) \ ,
\end{equation}
up to an irrelevant constant. The terms depending only on $v$, or only
on $v^\prime$, will drop out when inserted in the double integral
(\ref{minan}).  Setting finally $v-v^\prime = \sigma -\sigma^\prime$
(or $v-v^\prime = \sigma -\sigma^\prime -i\tau_0$) for points on the
same (or opposite) boundaries of the infinite strip, leads to the
expressions (\ref{vartn}) and (\ref{cas}), as claimed.
An alternative derivation of this result using the massless propagator
on the strip is 
\begin{eqnarray}\label{c12}
G_{11} (\sigma) &=& \sum_{n=-\infty}^{\infty} \int \frac{\rmd k}{2\pi}
\frac{\rme^{ik \sigma }}{k^{2} + (n\pi /\tau_0)^{2}} = \int \frac{\rmd
k}{2\pi}\frac{ \rme^{ik \sigma}}{k^{2}} + 2 \frac{\tau_0}{\pi}\sum_{n>0}
\frac{1}{n}\rme^{-n \pi |\sigma|/\tau_0} = -|\sigma| - \frac{2\tau_0}{\pi} \ln
\left(1-\rme^{-\pi |\sigma| /\tau_0} \right)\\
\label{c13} G_{12} (\sigma) &=& \sum_{n=-\infty}^{\infty} \int \frac{\rmd
k}{2\pi} \frac{(-1)^{n}\rme^{ik \sigma }}{k^{2} + (n\pi /\tau_0)^{2}}
= \int \frac{\rmd k}{2\pi}\frac{ \rme^{ik \sigma}}{k^{2}} + 2
\frac{\tau_0}{\pi}\sum_{n>0} \frac{(-1)^{n}}{n}\rme^{-n \pi
|\sigma|/\tau_0} = -|\sigma| - \frac{2\tau_0}{\pi} \ln
\left(1+\rme^{-\pi |\sigma| /\tau_0} \right)\qquad
\end{eqnarray}
These formulae agree with (\ref{cas}) up to an irrelevant constant.
\vskip 0.6 mm

It will be useful to write these expressions in Fourier space. This
can be done by using the identities
\begin{equation}\label{c5}
\sum_{n=-\infty}^\infty\, {1 \over b^2 + n^2} = {\pi\over b} \coth
(\pi b) \ , \ \ \ {\rm and}\ \ \ \ \ \sum_{n=-\infty}^\infty \, { \,
(-1)^n \over b^2 + n^2}= {\pi\over b \sinh (\pi b)} \ .
\end{equation}
To lighten the notation, we will suppress the label of the boundaries,
and use boldface letters for the corresponding vectors and matrices.
Thus ${\bf \tilde y}$ will stand for the two-component vector $(\tilde
y_1, \tilde y_2)$, and ${\bf G}$ for the $2\times 2$ matrix-valued
kernel $G_{J J^\prime}$.  With the help of the above formulae one
finds:
\begin{eqnarray}\label{c6}
{1\over 4\pi }\, \int _\sigma \hskip -1 mm \int_{\sigma^\prime} \,
{\rmd {\bf \tilde y}^{\, t} \over \rmd\sigma} \, {\bf G }(\sigma -
\sigma^\prime) \, {\rmd {\bf \tilde y}\over \rmd \sigma^\prime}
=\frac{1}{2} \int_k \tilde {\bf y}^{\, t}_k\, \widehat {\bf G}(k)\,
\tilde {\bf y}_{-k}\ \ \ ,
\end{eqnarray} 
where $t$ indicates the transpose of a vector, and
\begin{eqnarray}\label{c7} 
\widehat {\rm \bf G}(k) :=  k^2 \int
_{-\infty}^\infty {\rmd\sigma}\, \rme^{ik\sigma} \, {\rm \bf
G} (\sigma) = \left( \matrix { k \, {\rm coth} (\tau_0 k) \ \ &
\ \ - k / {\rm sinh} (\tau_0 k) \cr & \cr -k / {\rm sinh} (\tau_0 k) \
\ &\ \ k \, {\rm coth} (\tau_0 k) } \ \right) \ .
\end{eqnarray}
Since ${\rm det}\, \widehat {\bf G}(k) = k^2$, the inverse matrix
takes also a simple form: 
\begin{eqnarray}\label{c8} 
\widehat {\rm \bf G}(k)^{-1} \ \ =\ \, {1\over k}\, \left( \matrix {
\, {\rm coth} (\tau_0 k) \ \ & \ \ 1 / {\rm sinh} (\tau_0 k) \cr & \cr
1 / {\rm sinh} (\tau_0 k) \ \ &\ \ \, {\rm coth} (\tau_0 k) } \
\right) \ .
\end{eqnarray}
As a check note that in the limit of an infinitely-wide strip ($L\sim
\tau_0 \to\infty$) one finds $\widehat {\bf G}(k) \simeq \vert k\vert
\cdot \, {\bf 1}_{2\times 2}$. This is indeed the kernel for two
separate, half-infinite planes.  \vskip 0.5mm

Returning to the variational functional (\ref{vart1}), it can be
replaced by
\begin{eqnarray}\label{vartmat}
 \tilde {\cal A}({\bf a}, \tilde {\bf y}, \tilde {\bf z})\
= \ {1\over 2} \int _k \left( \, \tilde{\rm\bf y}_{k}^{\ t} \widehat
{\bf G}(k)\, \tilde {\rm\bf y}_{-k} + \tilde{\rm\bf z}_{k}^{\ t}
\widehat {\bf G}(k)\, \tilde {\rm\bf z}_{-k} \, \right) \, + \int_{k}
\left( {\bf a}_{k} \cdot {\bf H}_{-k} - \delta {\bf a}_{k} \cdot
\tilde {\bf z}_{-k}\right) \ , 
\end{eqnarray} 
where ${\bf a} \equiv (\alpha_1 , \alpha_2)$ is the vector of
Lagrange-multiplier fields, $\delta {\bf a} \equiv {\bf a}\, - \left<
{\bf a}\right> = (\alpha_1 -\cos\theta_0 , \alpha_2 + \cos\theta_0)$,
and ${\bf H}_k$ is the Fourier transform of the (vector of) composite
fields $h_J (\sigma +\tilde y_J(\sigma))$.  Solving the linear
equations for $\tilde {\bf z}$, and inserting the solution in the
above functional gives
\begin{equation}\label{vartvec} 
\tilde {\cal A}({\bf a},
\tilde {\bf y})\ = \ {1\over 2} \int _k \, \tilde{\rm\bf y}_{k}\,
\widehat {\bf G}(k)\, \tilde {\rm\bf y}_{-k} - {1\over 2} \int _k \,
\delta {\bf a}_{k}\, \widehat {\bf G}(k)^{-1} \delta {\bf a}_{-k} \,
\, + \int_{k} {\bf a}_{k} \cdot {\bf H}_{-k} \ .  
\end{equation}
We can now read off the Feynman rules that generalize the ones of the
previous section.  The propagators and 1-point functions for the
vector fields are:
\begin{eqnarray}\label{a3n} \diagram{dcostheta}\ &:= &\ \left<\, {\bf
a}_k \,
 \right>\ =   \ (\cos \theta_0 ,  -\cos\theta_0)\,   \delta(k) \nonumber \\
  \diagram{line}\ &:=&\ \langle \delta {\bf a}_k\, \delta {\bf
a}_{-k}^{\, t}\rangle \ =
  \ \widehat {\bf G}(k)  \\
 \diagram{wiggly} \ &:=&\ \left< \tilde {\bf y}_k\, \tilde {\bf y}^{\,
t}_{-k}\right>\ =\ - \widehat{\bf G}(k)^{-1} \ .  \nonumber
\end{eqnarray} 
The vertices do not mix fields on opposite boundaries, and are thus
two copies of the vertices in (\ref{a4}).  \vskip 0.9 mm

Using these rules we may calculate the energy to any desired order in
the $h$-expansion.  The leading, quadratic energy that generalizes the
JdG result reads:
\begin{eqnarray}\label{c9}  
E_2^{\rm strip}  &  =& 
\gamma\, \Biggl(  \diagram{dwig0}+ \diagram{ord2a} \Biggr)\ \nonumber\\
& =& \gamma \sin^2\theta_0 \, \int_0^{\infty} \frac{\rmd k}{2\pi } \,
k\, \Biggl( \left( \vert h_{1,k}\vert^2 + \vert h_{2,k}\vert^2 \right)
\, {\cosh (kL/\sin\theta_0) -1 \over \sinh (kL/\sin\theta_0) }\ + \
\vert h_{1,k} - h_{2,k}\vert^2 / \sinh (kL/\sin\theta_0) \Biggr) \
.\qquad
\end{eqnarray}
Since both terms inside the integral are positive-definite, 
it is energetically favorable for $h_{1,k}$ and $h_{2,k}$ to have the same phase. Thus
the interaction between the two contact lines is attractive.  Note
that if we fix $h_{1}$ and allow $h_{2}$ to freely adjust, we find that the
minimum of the energy is obtained for
\begin{equation}\label{c10}
h_{2} (k) = \frac{h_{1} (k)}{\cosh (kL/\sin \theta_{0})}\ .
\end{equation}
The energy for given $h_{1}$ and free $h_{2}$ thus reads 
\begin{equation}\label{c11}
E_2^{\rm strip} \Big|_{\mathrm{free~} h_{2}} =\gamma \, \sin^2
\theta_{0} \int_{0}^{\infty} \frac{\rmd k}{2\pi} \, k\,
|h_{1,k}|^{2}\, \tanh \!\left(\frac{kL}{\sin \theta_{0}} \right)\ .
\end{equation}
In the limit of $L\to \infty$, we recover our previous expression (\ref{JDG}) as expected.

Taking the same limit in (\ref{c9}) shows that the
interaction decays exponentially, as $\sim {\rm exp}
(-2kL/\sin\theta_0)$.  This exponential decay also applies for fixed
$L$ and very small contact angle, since the actual separation of the
(unperturbed) contact lines is $L/\sin\theta_0$.  In the opposite
limit of a thin strip, or equivalently of very long-wavelength
deformations, we find:
\begin{equation}\label{94} 
E_2^{\rm strip} \, \simeq \, \gamma  \sin\theta_0 L\,
\int_0^{\infty} \frac{\rmd k}{2\pi}\, \Bigl[\, \vert h_{1,k} -
h_{2,k}\vert^2 \, \left({\sin^2\theta_0\over L^2} - {k^2\over
6}\right) \ + \ \frac{k^2}2 \left( \vert h_{1,k}\vert^2 + \vert h_{2,k}\vert^2
\right) \ +\ O(k^4) \, \Bigr] \ \ .
\end{equation}
The leading term has a simple geometric interpretation: It is
proportional to the increase in area of a planar strip, whose
boundaries undergo a relative displacement $h_1-h_2$ along the walls,
with which it made initially an angle $\theta_0$.  For $h_1=h_2$, the
next term in the above quadratic energy corresponds to an elastic rod
with effective tension $\gamma_{\rm eff} = \gamma L\sin\theta_0$. This
has also a simple geometric interpretation: The rod is in fact a thin
strip, of width $L/\sin\theta_0$, which is deformed by an amount\
$h_1(y) \sin\theta_0$ in the transverse direction.
  
%\vskip 3mm
  
%%%%%%%%%%%%%%%%%%%%%%%%%%%%%%%%%%%%%%%%%
 
\section{Discussion}
\label{s:8}

In the previous sections we have shown how to calculate the energy of
a deformed, almost rectilinear, contact line to any desired order in
perturbation theory.  We would now like to discuss some general
properties of this expansion.  One important point is that
perturbation theory is {\it quasi-local}, i.e.\ the total energy is
concentrated in a region of size equal to the typical wavelength of
the deformation.  We indeed saw that, as long as the large-volume
cutoff has been fine-tuned so as to cancel the global tadpole, it
decouples from any localized perturbation.  One would expect the same
to be true for all other geometric length scales of the system, such
as the wall's inverse curvature.  If this is true, at sufficiently
short distances perturbation theory should be scale-covariant, as was
pointed out in section \ref{s:5}.  In momentum space, the scaling
symmetry (\ref{rescale}) reads:
\begin{equation}\label{virt}
 \tilde E[h^{(\lambda)}]\, = \ \lambda^2 \, \tilde E[h]\ \ \ \ \ \
{\rm for}\ \ \ \ \ \ \ h^{(\lambda)}_{\, k} = \lambda^2 \, h_{\lambda
k}\ .
\end{equation}
Inspection of eqs.~(\ref{a6})--(\ref{endd}) shows that this indeed
holds at each order of the expansion, and even for each
individual diagram.  Note, in passing, that the scaling symmetry does
not imply conformal invariance, as would have been the case if the
one-dimensional theory were truly local.  \vskip 0.5mm

Finiteness of the JdG quadratic energy requires that
\begin{equation}\label{falloff} k h_k \to 0\ \ , \ \ \ {\rm for\
both}\ \ \ \ \ k\to \infty \ \ \ {\rm and}\ \ \ \ k\to 0 \ .
\end{equation} 
In other words, $h(y)$ must be continuous everywhere and finite, and
it must vanish as $y\to\pm\infty$.  A more stringent condition is, in
fact, required to prove ultraviolet finiteness at all higher orders.
It reads
\begin{equation}\label{fallo} k^2 h_k \to 0\ \ \ \ \ {\rm for}\ \ \ \
\ \ k\to \infty\ \ \ \ \ \Longleftrightarrow \ \ \ \ h^{(\lambda)}_k
\to 0 \ \ \ \ \ {\rm for}\ \ \ \lambda\to\infty\ .
\end{equation} 
Stated differently, the profile function $h(y)$ must also have a
continuous first derivative.  That this is indeed necessary follows by
considering for instance the ``comb'' diagrams, the first few being
(\ref{a10}), (\ref{e6n}) and (\ref{e8n}).  As the reader can check, a
power fall-off slower than (\ref{fallo}) would make the comb diagrams
with a large enough number of vertices diverge.  To show that this
condition is also sufficient, it is convenient to assign the scaling
dimensions $[k]= 1$ and $[h_k] = -2$ to the factors entering in a
diagram.  Because $k^2h_k\to 0$ at high momentum, the degree of
divergence of any partial integration is bounded from above by the
corresponding scaling dimension, in which one only counts elements
that depend on the integrated momenta.  The scaling symmetry
(\ref{virt}) implies that the overall scaling dimension of any tree
diagram is $-2$, so there is no divergence from the integration region
where all the momenta go to infinity.  Keeping one (or more) of the
momenta fixed amounts to removing from the counting a factor $\rmd k\,
k^m\, h_k$, and at most one solid and $m$ curly propagators that
emanate from the corresponding vertex.  This can only lower the
scaling dimension, so all the partial integrations are also
ultraviolet finite. {\it q.e.d}. 

\vskip 0.5mm Infrared finiteness is
trickier to establish diagrammatically. Condition (\ref{falloff})
suffices to ensure that there is no divergence when the momenta
flowing into individual vertices go to zero.  The dangerous diagrams
are, however, those for which such momenta {\it add up} to zero along
some curly line. Inspection of the expression (\ref{inspect}) shows,
nevertheless, that the result is finite up to quartic order, thanks to
the Heaviside functions that multiply such dangerous terms.  To prove
finiteness at all higher orders, it is more convenient to go back to
the pair of classical equations (\ref{bc1n}) and (\ref{constr}).  Let
$\tilde y^{(n)}(\sigma)$ and $\tilde z^{(n)}(\sigma)$ be the solutions
of these equations at $n$th order. It is then straightforward to check
that, if these functions vanish at $y\to\pm\infty$ for all $n\leq N$,
they will continue to do so for $n=N+1$.  This is in turn sufficient
to guarantee the infrared finiteness of the energy at all orders.
\vskip 0.5mm

What about non-perturbative effects?  To fix ideas, let $h(y) =
\varepsilon\, f(y)$ with $f(y)$ a given profile function, and
$\varepsilon$ the parameter controlling the perturbative expansion.
One expects that the radius of convergence of this expansion is
finite, since at large enough $\varepsilon$ the solution to
eqs.~(\ref{bc1n}) and (\ref{constr}) should stop being analytic.  This
could signal either one of the following two things: (i) that our
parameterization is singular, or (ii) that the surface develops real
geometric singularities or that there is a change in topology.  It
would be very interesting to find some general criteria which could
distinguish between these two possibilities.  Note that a topological
transition may occur if it is energetically favorable to drill two
holes in the fluid surface, and to replace the corresponding disks by
a cylinder.  In any case, the following simple (though rather crude)
linear bound
\begin{equation} \tilde E[h]\, < \, \gamma \int_y \, \vert
h\vert \, +\, \Bigl\vert \gamma^\prime \int_y\, h\, \Bigr\vert
\end{equation} 
shows that the energy of a pinned contact line stays
finite. Furthermore, localized microscopic perturbations always have a
vanishingly-small energy, and should decouple from the physics at
longer scales.  \vskip 0.5mm

This brings us to our final remark \cite{green60}: as was explained in
section \ref{s:2}, the purely-Dirichlet minimal surface problem is
related to the mixed Dirichlet-Neumann problem, relevant for capillary
phenomena, by a Legendre transformation. A Legendre transformation
looks at first sight rather benign, but it drastically modifies the
nature of perturbation theory.  This is best illustrated by the
following spectacular phenomenon \cite{Finn2}: A wedge in the tubular
contour $\partial\Omega$ of section \ref{s:2}, with opening angle less
than $\vert \pi - 2\theta_0\vert$, is a local geometric obstruction
which forces the capillary surface to develop a second sheet.  This
has been observed in micro-gravity experiments.  Notice that the wedge
can be of microscopic transverse size, but it should extend to all
values of the height coordinate $z$.  It is the latter assumption
which is responsible for the apparent non-decoupling of short-distance
scales.  The story is reminiscent of the role played by wormholes in
theories of quantum gravity.  This analogy, as well as the possible
impact of wedge defects on the problem of wetting, deserve further
investigation.

%%%%%%%%%%%%%%%%%%%%%%%%%%%%%%%%%%%%%%%%
 \vskip 1cm 

{\bf Acknowledgments}: We thank Angelina Aessopos, Denis Bernard,
Fran\c cois David, Matthias Gaberdiel, Gary Gibbons, Ramin
Golestanian, Gian Michele Graf, Yves Pomeau, Elie Rapha\"el and
Konstantin Zarembo for useful conversations.  This research was
partially supported by the European Networks MRTN-CT-2004-512194, 
MRTN-CT-2004-005104 and ANR program 05-BLAN-0099-01.

%%%%%%%%%%%%%%%%%%%%%%%%%%%%%%%%%%%%%%%%

%\vskip 0.6cm 

 \appendix

\section{Weierstrass parameterization} 
\label{s:Weierstrass}
\begin{figure}[b]
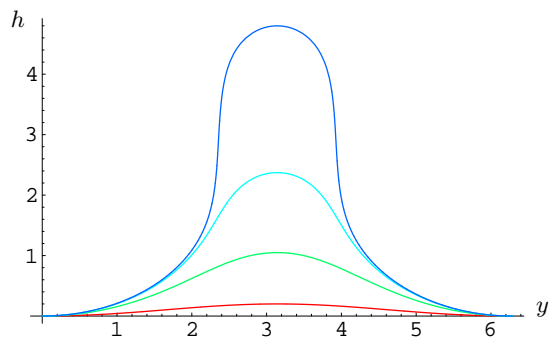
 $\rule{0mm}{4.3cm}^{\displaystyle
h}\!\!\!\!\fig{7cm}{weierstrass1} {\rule{0mm}{0.5cm}^{\displaystyle
y}}$ \caption{Parametric plot of (\ref{A4}) for $\kappa
=0.1,0.5,0.9,0.999$. Increasing $\kappa$ increases the amplitude of $h
(y)$, and leads to a singularity at $\kappa =1$. The function $\tilde
Z (w)$ has been shifted s.t.\ $h (0)=0$. } \label{f:weierstrass}
\end{figure}
In section \ref{s:4} we have parameterized the minimal surfaces in
terms of two functions $\tilde Y(w)$ and $\tilde Z(w)$, which are
related by the conformal-gauge condition (\ref{constr}).  The
parameterization is global provided the two functions are analytic
everywhere in the upper-half complex plane.  This is indeed the case
in perturbation theory, but more generally, for a given analytic
$\tilde Z$, the solution of (\ref{constr}) will not give an analytic
function $\tilde Y$.  A constructive solution of the conformal-gauge
condition, that guarantees analyticity, is given by the Weierstrass
representation
{\renewcommand{\arraystretch}{1.2}
\begin{equation}\label{wei}
\left( 
\begin{array}{c}
X\\
Y\\
Z
\end{array} \right)(w)  =  \int_{0}^w  \rmd v\, f \times\,
\left(
\begin{array}{c}
 2g \\
 -i (1+g^2)\\
 1 - g^2 
\end{array} \right)
\ \ ,
\end{equation}
where $f(v)$ and $g(v)$ are holomorphic functions in the upper-half
plane.  To go to the special gauge eq.~(\ref{choice}) of section
\ref{s:4}, one sets $2f = ic/g$. The surfaces are then parameterized by
a single function: 
\begin{equation}\label{}
 \left(\begin{array}{c} 
X \\
Y\\
Z 
\end{array}
 \right)(w)  =  {ic \over 2}\, \int_{0}^w \rmd v\,
\left( 
\begin{array}{c}
 -2 \\
 i (g + {1/ g})\\
 g - {1/g} 
\end{array}
\right) \ \ .
\end{equation}}
Clearly, this special parameterization is non-singular if and only if
$g$ has no zeroes in the upper-half complex plane.  Since in the
expression (\ref{wei}) both $f$ and $g$ are allowed to have any number
of zeroes, this shows that the condition (\ref{choice}) need not
always define a good global gauge.  \vskip 0.5mm

To describe the deformed surfaces of section \ref{s:4} we write
\begin{eqnarray}
g = g_0 + \tilde g\ , \ \ \ \ {\rm with}\ \ \ \ \ \ g_0 = \
{\cos\theta_0 - 1\over \sin\theta_0}\ .
\end{eqnarray}
The unperturbed planar surface corresponds to $\tilde g =0$. Other
choices of $\tilde g$, which are holomorphic in the upper-half plane
(including the point at infinity) and for which $g_0 + \tilde g$ has
no zeroes, describe globally-parameterized deformed fluid surfaces. As
a simple example, let $\theta_0=\pi/2$ and take
\begin{equation}\label{A4}
\renewcommand{\arraystretch}{2.4}
g (v)= -1 - \kappa\, \rme^{2 i v}\qquad \Longrightarrow
\qquad 
\begin{array}{l}
\displaystyle 
  \tilde Y
(w) = \frac{i}{4} \left[ -\kappa \rme^{2 i w}  +\log
\left({1+\kappa\, \rme^{2 i w} } \right)\right]\\
\displaystyle 
\tilde Z (w)= -\frac{1}{4} \left[ \kappa\, \rme^{2 i w} + \log
\left({1+\kappa\, \rme^{2 i w} } \right)\right]
\end{array}
\end{equation}
 \noindent where $\kappa$ is a real parameter between 0 and 1, and in
the expressions for $\tilde Y$ and $\tilde Z$ we have dropped an
irrelevant constant [which can be absorbed in a redefinition of the
origin of coordinates].  For small $\kappa$, this function describes a
periodic minimal surface with period $\Delta y = 2\pi$, and with a
deformed contact-line given by $h(y) = \kappa \cos y + O(\kappa^2)$.
For $\kappa$ finite, the contact-line profile is a complicated
function given implicitly by eqs.~(\ref{A4}), and plotted on figure
\ref{f:weierstrass}.  Inserting the above $\tilde Y$ and $\tilde Z$ in
the expression (\ref{enfina}) for the energy gives:
\begin{equation}
\tilde E / {\rm period} = \, {\pi\gamma\over 4} \left[ \kappa^2 - {\rm
log} ( 1 - \kappa^2) \right] \ .
\end{equation} 
This reduces to the JdG energy at small $\kappa$, and can also be
verified numerically.  Note that when $\kappa \to 1$ the surface
becomes singular, and the energy per period diverges.

%{\bf Kay:  Peux tu verifier a quelle valuer de
%$\kappa$ l'algorithme numerique se casse la figure?
%KAY: La divergence est differente.}

%\vskip 0.6cm 
%%%%%%%%%%%%%%%%%%%%%%%%%%%%%%%%%%%%%%%%%%

\section{More general large-volume cutoff}
\label{s:B}
In this appendix we will repeat the calculation of the energy of
section \ref{s:4}, using a more general container with an outer wall
at an arbitrary inclination angle. The characteristic function
$\Theta_C(\vec r)$ now reads \begin{equation}\label{cont}
\Theta_C(\vec r)\ =\ \Theta(x) \, \Theta \!\left(y+{L_y\over 2}\right)
\Theta\!\left({L_y\over 2}-y\right) \Theta(L_x - x \cos\phi + z\sin\phi )\ .
\end{equation}
The inclination angle $\phi$ of the outer wall is a control parameter,
which should drop out in the $L_x, L_y\to\infty$ limit. The contact
angle of the planar surface with this outer wall is equal to
$\pi-\phi-\theta_{0}$, so Young's equilibrium condition requires that
the corresponding solid-fluid tension be $\gamma^{\prime\prime} =
-\cos (\phi+\theta_{0})$.  Repeating the same steps as in section
\ref{s:4} leads to the general expression for the energy
\begin{eqnarray}\label{b2}
\tilde{E}[h]\ = \ - {\gamma\over 2} \oint_{\partial D} \Delta\vec r
\cdot \partial_\perp \Delta \vec r \ - {\gamma} \oint_{\partial D}
\Delta\vec r \cdot \partial_\perp \vec r_0 \ + {\gamma}
\oint_{\partial{\cal D}} \vert \hat n\cdot \partial_\perp \vec
r_0\vert^{-1}\ \hat n \cdot \Delta\vec r \ \ + \ {\cal E}_{\rm bnry} \
\ ,
\end{eqnarray}
\noindent where ${\cal D} = [0, \tau_0]\times [-L_y/2 , L_y/2]$ is the
parameter domain defined by $\Theta_C(\vec r_0)=1$, \ $\partial_\perp$
is the derivative in the inward normal direction to $\partial {\cal
D}$, and $\hat n$ is the three-dimensional vector normal to the
container boundary.  \vskip 0.5mm

We can now verify that the inclined wall does not contribute to the
above expression. This follows from a fine cancellation between the
three last terms in eq.~(\ref{b2}):
\begin{equation}\label{b3} 
-\ \gamma\cos\theta_{0} \ \int_{\sigma}
\tilde z \ + \ {\gamma \sin\phi\over \sin(\phi +\theta_{0})}
\int_{\sigma} \tilde z \ -\ {\gamma^{\prime\prime}\sin\theta_{0} \over
\sin(\phi+ \theta_{0})}\, \int_{\sigma} \tilde z \ \ \
\Biggr\vert_{\tau = \tau_0} \ = \ 0\ \ .
\end{equation}
We here used the normal vector  $\hat n = (-\cos\phi, 0, \sin\phi)$,
which implies that \ $\vert \hat n\cdot \partial_\perp \vec r_0\vert =
\sin(\phi +\theta_{0})$, as well as some three-dimensional geometry
which is required to extract the contribution of ${\cal E}_{\rm
bnry}$.  Doing some straightforward trigonometry, and using the fact
that $\gamma^{\prime\prime} =\cos(\phi +\theta_{0})$, one can check
that the three terms (\ref{b3}) indeed cancel.  This confirms the
decoupling of the large-volume cutoff, as was announced in section
\ref{s:4}.
  
%\vskip 1cm 

%\vskip 3cm

%\bibliography{citation}

\begin{thebibliography}{10}

 
\bibitem{math} See for instance J.C.C. Nitsche, {\it Introduction to
minimal surfaces}, Cambridge U. Press (Cambridge 1989); U. Dierkes,
S. Hildebrandt, A. K\"uster and O. Wohlrab, {\it Minimal surfaces},
vols. I and II , Springer-Verlag (Berlin 1991); R. Osserman, editor,
{\it Minimal surfaces}, Springer-Verlag (Berlin 1997), and references
therein.

\bibitem{Plateau} J.A.F. Plateau, {\it Statique exp\'erimentale et
th\'eorique des liquides soumis aux seules forces mol\'eculaires},
Clemm (Paris 1873).

\bibitem{Isenberg} C.~Isenberg, {\it The science of soap films and
soap bubbles}, Dover (New York 1992).

\bibitem{boudaoud} see e.g.\ A. Boudaoud, P. Patricio and M. Ben Amar,
{\it The helicoid versus the catenoid: Geometrically induced
bifurcations}, Phys. Rev. Lett.  {\bf 83}, 3836 (1999) and references
therein.

\bibitem{bp} D.~J.~Gross and P.~F.~Mende, {\it The high-energy
behavior of string scattering amplitudes},
  Phys.\ Lett.\ B {\bf 197}, 129 (1987);  \\
C.~Bachas and B.~Pioline, {\it High-energy scattering on distant
branes}, JHEP {\bf 9912}, 004 (1999) [arXiv:hep-th/9909171].

\bibitem{Gennes}
P.G. de Gennes,  {\it Wetting: statics and dynamics},
Rev. Mod. Phys. {\bf 57},  827 (1985);\\
P.G. de Gennes, F. Brochard-Wyart and D. Quer\'e, {\it Capillarity and
wetting phenomena: drops, bubbles, pearls and waves}, Springer (New
York 2004).
 
\bibitem{Young} T.~Young, {\it An essay on the cohesion of fluids},
Phil. Trans. Roy. Soc. London {\bf 95}, 65-87 (1805).

\bibitem{Laplace} P.-S. Laplace, {\it Trait\'e de m\'ecanique
c\'eleste}, Courcier (Paris 1805).
 
\bibitem{helein} F.~H\'elein, {\it Constant mean curvature surfaces,
harmonic maps and integrable systems}, Birkh\"auser (Basel 2000).


%%%%%from bibtex

\bibitem{JoannyDeGennes1984} J.F. Joanny and P.G.~de Gennes, \newblock
{\em A model for contact-angle hysteresis}, \newblock
J. Chem. Phys. {\bf 81}, 552 (1984); see also\\ Y. Pomeau and
J. Vannimenus, {\it Contact angle on heterogeneous surfaces: weak
heterogeneities}, Journal of Colloid and Interface Science {\bf 104},
477 (1985).

\bibitem{japonais} see for instance K.~Sekimoto, R.~Oguma and
K.~Kawasaki, {\it Morphological stability analysis of partial
wetting}, Ann. Phys.  {\bf 176}, 359 (1987), and references therein.

\bibitem{GolestanianRaphael2001} R.~Golestanian and E.~Rapha\"el,
\newblock {\em Relaxation of a moving contact line and the
Landau-Levich effect}, \newblock Europhys. Lett. {\bf 55}, 228 (2001)
[arXiv:cond-mat/0006496]; erratum, Europhys. Lett. {\bf 57}, 304
(2002); R.~Golestanian and E.~Rapha\"el, {\it Roughening transition in
a moving contact line}, Phys. Rev. E {\bf 67}, 031603 (2003)
[arXiv:cond-mat/0204531].
 
\bibitem{LeDoussalWieseRaphaelGolestanian2004} P.~Le Doussal, K.J.\
Wiese, E.~Rapha\"el and R.~Golestanian, \newblock {\em Can non-linear
elasticity explain contact-line roughness at depinning?},
Phys. Rev. Lett. {\bf 96}, 015702 (2006) \newblock
[arXiv:cond-mat\slash {0411652}] and to be published. 
 
\bibitem{PrevostRolleyGuthmann2002} A.~Prevost, E.~Rolley and
C.~Guthmann, \newblock {\em Dynamics of a helium-4 meniscus on a
strongly disordered cesium substrate}, \newblock Phys. Rev. B {\bf
65}, 064517 (2002); S.~Moulinet, C.~Guthmann and E.~Rolley, {\it
Roughness and dynamics of a contact line of a viscous fluid on a
disordered substrate}, Eur. Phys. J. E {\bf 8}, 437 (2002).

\bibitem{LeDoussalWieseChauve2002} P.~Le Doussal, K.J. Wiese and
P.~Chauve, \newblock {\em 2-loop functional renormalization group
analysis of the depinning transition}, \newblock Phys. Rev. B {\bf
66}, 174201 (2002) \newblock [arXiv:cond-mat/{0205108}].

\bibitem{RossoKrauth2002} A.~Rosso and W.~Krauth, \newblock {\em
Roughness at the depinning threshold for a long-range elastic string},
\newblock Phys. Rev. E {\bf 65}, 025101 (2002).

\bibitem{MoulinetRossoKrauthRolley2004} S.~Moulinet, A.~Rosso,
W.~Krauth and E.~Rolley, \newblock {\em Width distribution of contact
lines on a disordered substrate}, \newblock Phys. Rev. E {\bf 69},
035103 (2004) \newblock [arXiv:cond-mat/0310173].


\bibitem{Pomeau} see e.g.\ M. Ben Amar, L.J. Cummings and Y. Pomeau,
{\it Transition of a moving contact line from smooth to angular},
Physics of Fluids {\bf 15}, 2949 (2003) and references therein.

\bibitem{Finn} E. Giusti, {\it Minimal Surfaces and Functions of
Bounded Variation}, Birkh\"auser-Verlag (Basel 1984); R.~Finn, {\it
Equilibrium Capillary Surfaces}, Springer-Verlag (new York 1986).

\bibitem{Caldeira} A.~O.~Caldeira and A.~J.~Leggett, {\it Influence Of
Dissipation On Quantum Tunneling In Macroscopic Systems,} Phys.\ Rev.\
Lett.\ {\bf 46} (1981) 211; {\it Quantum Tunneling In A Dissipative
System,} Annals Phys.\ {\bf 149} (1983) 374.
 

\bibitem{Callan} C.~G.~.~Callan and L.~Thorlacius, {\it Open string
theory as dissipative quantum mechanics}, Nucl.\ Phys.\ B {\bf 329},
117 (1990).
 
\bibitem{LeDoussalGolestanian} P.~Le Doussal and R.~Golestanian, unpublished. 

\bibitem{david} For a pedagogical reference see e.g.\ F.~David, {\it
Geometry and field theory of random surfaces and membranes,} in
proceedings of the 5th Jerusalem school, D. Nelson et al editors,
World Scientific (Singapore 1989).
 
\bibitem{compl} see e.g.\ R. Churchill, J. Brown and R. Verhey, {\it
Complex Variables and Applications}, McGraw-Hill (New York 1975).
 
\bibitem{bg} C.~Bachas, {\it Relativistic string in a pulse,} Annals
Phys.\ {\bf 305}, 286 (2003) [arXiv:hep-th/0212217]; C.~P.~Bachas and
M.~R.~Gaberdiel, {\it World-sheet duality for D-branes with travelling
waves,} JHEP {\bf 0403}, 015 (2004) [arXiv:hep-th/0310017].

\bibitem{green60} C. Bachas, talk at Michael Green's 60th birthday,
{http://www.damtp.cam.ac.uk/eurostrings06}.
 
\bibitem{Finn2} R.~Finn, {\it Capillary Surface Interfaces,} Notices
of the AMS vol. {\bf 46}, no. 7, and references therein.
 
 
\end{thebibliography}
%\bibliographystyle{../../macros/KAY}
%\bibstyle{../macros/KAY}
%\end{document} 

\end{document}